\documentclass[useAMS,usenatbib]{mnras}

\usepackage{epsfig,color,graphicx,subfigure}
\usepackage{rotating}
\usepackage{longtable} % for 'longtable' environment
\usepackage{pdflscape}
\usepackage{siunitx}

\newcommand{\etal}{et~al.} 
\newcommand{\ionhy}{H{\sc ii} }

\newcommand{\kms}{$\mbox{km~s}^{-1}$ }
\newcommand{\kmsns}{$\mbox{km~s}^{-1}$}

\usepackage{lipsum}
\usepackage{array,booktabs,dcolumn,calc}
\usepackage[fleqn]{amsmath}
\usepackage{float}
\usepackage{bm}
\usepackage{amssymb,amsmath}
\usepackage{setspace}
\usepackage{rotating}
\usepackage{xcolor}

\bibpunct{(}{)}{;}{a}{}{,}

\def\apj{{ApJ}}                 
\def\apjl{{ApJ}}                
\def\apjs{{ApJS}}

\def\aap{{A\&A}}

\def\mnras{{MNRAS}}

\def\pasp{{PASP}} 
\def\pasa{{PASA}}

\date{Accepted 2018 November 15. Received 2018 November 14; in original form 2018 October 2.}

\begin{document}

\title[Star forming clumps in Carina]{Environmental conditions shaping star formation: The Carina Nebula}
\author[Y. Contreras \etal]{Y. Contreras,$^{1}$\thanks{Email: ycontreras@strw.leidenuniv.nl} 
D. Rebolledo,$^{2,5}$ 
S.\ L. Breen,$^{3}$ 
A.\ J. Green,$^3$ 
M.\ G. Burton,$^4$ 
\\
$^1$ Leiden Observatory, Leiden University, P.O. Box 9513, NL-2300 RA Leiden, The Netherlands;\\
$^2$Joint ALMA Observatory, Alonso de C\'ordova 3107, Vitacura, Santiago, Chile.\\
$^3$ Sydney Institute for Astronomy (SIfA), School of Physics, University of Sydney, NSW 2006, Australia;\\
$^4$Armagh Observatory and Planetarium, College Hill, Armagh, BT61 9DG, Northern Ireland, UK\\
$^5$National Radio Astronomy Observatory, 520 Edgemont Road, Charlottesville, VA 22903, USA
}

 \maketitle
  
 \begin{abstract}

Using the Mopra telescope, we have targeted 61 regions in the Carina Nebula, covering an area of 1.5 square degrees, of bright and compact 870 $\mu$m dust continuum emission for molecular line emission from a host of 16 spectral lines at 3mm, including several dense gas tracers. We found that the clumps detected in Carina in general have in average higher temperatures (27 K compared to 21 K), and lower masses (214 M$_\odot$ compared to 508 M$_\odot$) than clumps located at a similar distance to us in the Galactic Plane. We compare the properties of the molecular line emission of these clumps with the MALT90 survey, finding that the detection rates of the molecular lines are similar to MALT90 clumps that are classified as PDRs. However, most of the clumps located within 10$\arcmin$ of $\eta$ Carina have little molecular line emission detected in our observations. Given the lack of maser detection in the Carina region, we also compared the properties the clumps in Carina to those of Galactic clumps associated with 6.7-GHz methanol masers. We found that the clumps in Carina are warmer, less massive, and show less emission from the four most commonly detected molecules, HCO$^+$, N$_2$H$^+$, HCN, and HNC, compared to clumps associated with masers in the Galactic Plane. Overall our results are consistent with the scenario in which the high radiation field of $\eta$ Carina is dramatically affecting its local environment, and therefore the chemical composition of the dense clumps.

\end{abstract}


\section{Introduction}

The Carina Nebula (NGC 3372) is a giant and rich star-forming region located in the southern Galactic plane \citep[see][]{Smith-2007}. It hosts more than 60 massive O and B stars including $\eta$ Carina and several hundred protostars \citep{Smith-2010} making it one of the most spectacular regions of high-mass star formation in the Galaxy. The Carina Nebula (hereafter Carina) is located relatively nearby \citep[distance 2.3$\pm$0.1 kpc,][]{Smith-2002} and  our line of sight toward Carina is not obscured by intervening Galactic extinction, making it an excellent region to study star formation, feedback processes and triggered star formation. Moreover, since Carina has an unusually high concentration of high-mass stars its provides an unique laboratory to study the effects that these high-mass stars have on their surrounding interstellar medium (ISM). 

Observations with the \textit{Hubble} \citep{Smith-2010b} and \textit{Spitzer} \citep{Smith-2010} telescopes, have shown the presence of dust pillars in Carina. These dust pillars correspond to the regions where star formation is currently ongoing, containing several hundreds of young proto-stellar objects (ages $\sim$10$^5$ years). In total, Carina contains more than 60000 young stars \citep{preibisch-2011}, and the star formation rate of this region of 0.017 M$_\odot$ yr$^{-1}$, which contributes to 1\% of the star formation rate of the Galaxy, making it one of the most important star forming regions in our Galaxy \citep{Povich-2011,preibisch-2011}. 

Despite having a high-concentration of high-mass stars, no high-mass young-stellar objects (YSOs, $>20$ M$_\odot$) have been found in Carina, but rather a population of low to intermediate mass YSOs ($1<$ M $<10$ M$_\odot$), with low luminosities ($<10^4$ L$_\odot$) located at the edges of the Nebula \citep{Gaczkowski-2013}. 

Sub-millimeter observations with the Large APEX Bolometer Camera (LABOCA) from the Atacama Pathfinder EXperiment (APEX) telescope of the dense ($>10^{21}$ cm$^{-2}$), cold ($<20$ K) gas have shown that dense molecular clouds still exist close to the OB stars in this region. However the total dust mass in dense clouds only account for 10\% of the total mass budget of Carina \citep{preibisch-2011}.

Recently, Carina has become the target of a number of observations at radio wavelengths, revealing the large-scale view of the molecular complex through comprehensive observations of $^{12}$CO and $^{13}$CO \citep{Rebolledo-2016}, as well as H{\sc i} \citep{Rebolledo-2017}. These observations have shown that the fraction of the molecular gas changes across the complex, but it can achieve values $\sim$ 80\% of the total mass in some regions\citep{Rebolledo-2016}.

Here we present Mopra observations of the 61 densest clumps within Carina, which have been revealed by LABOCA 870 $\mu$m dust continuum observations of the ATLASGAL survey \citep[APEX Telescope Large Area Survey of the Galaxy;][]{Schuller-2009}. In order to determine the physical and chemical evolutionary stages of the 61 densest clumps, we have performed observations of the dense molecular gas with the Mopra telescope. To compare the chemical and physical properties of these clumps to other star forming regions in the Galaxy, we used in our observations the same frequency setup as the MALT90 survey \citep[Millimetre Astronomy Legacy Team 90 GHz;][]{Foster-2013,jackson-2013,Rathborne-2016}. The MALT90 survey characterized the physical and chemical evolution of 3246 high-mass star-forming clumps in the inner Galaxy, by observing their molecular line emission in 16 spectral lines. This survey also observed clumps in a range of evolutionary stages, from quiescent pre-stellar clump where star formation has not begun yet, to proto-stellar clumps, having embedded proto-stars, to more evolved HII Regions and clumps associated with Photodisociation Regions (PDRs) \citep[see ][ for a detailed description of the clump classification]{jackson-2013}.

\section{Observations and data reduction}

\subsection{The targets}

We targeted 61 dense clumps that were selected based on their dust continuum emission at 870 $\mu$m from  the ATLASGAL survey \citep{Schuller-2009,Contreras-2013a,Urquhart-2014}. The ATLASGAL survey has produced a complete census of dense high-mass clumps over a range of evolutionary stages, ranging from quiescent pre-stellar clumps, showing no emission at infrared wavelengths to those that contain proto-stars and more evolved clumps associated with \ionhy regions and PDRs. In Carina, ATLASGAL detected 91 870 $\mu$m compact clumps, of which we targeted the brightest 61 with Mopra, in order to analyse the properties of the accompanying molecular line emission.

The 870 $\mu$m flux of of the 61 target clumps range from 0.47 to 4.53 Jy Beam$^{-1}$. Their angular sizes range from 9 to 166\arcsec, which at the distance of the Carina Nebula  \citep[2.3 kpc;][]{Smith-2010b} corresponds to physical sizes of 0.1 to 1.9 pc, with a mean value of 0.6 pc.

\begin{figure*}
   	\centering
   	\includegraphics[trim={3cm 2cm 2cm 6cm}, angle=-90, clip,width=0.9\textwidth]{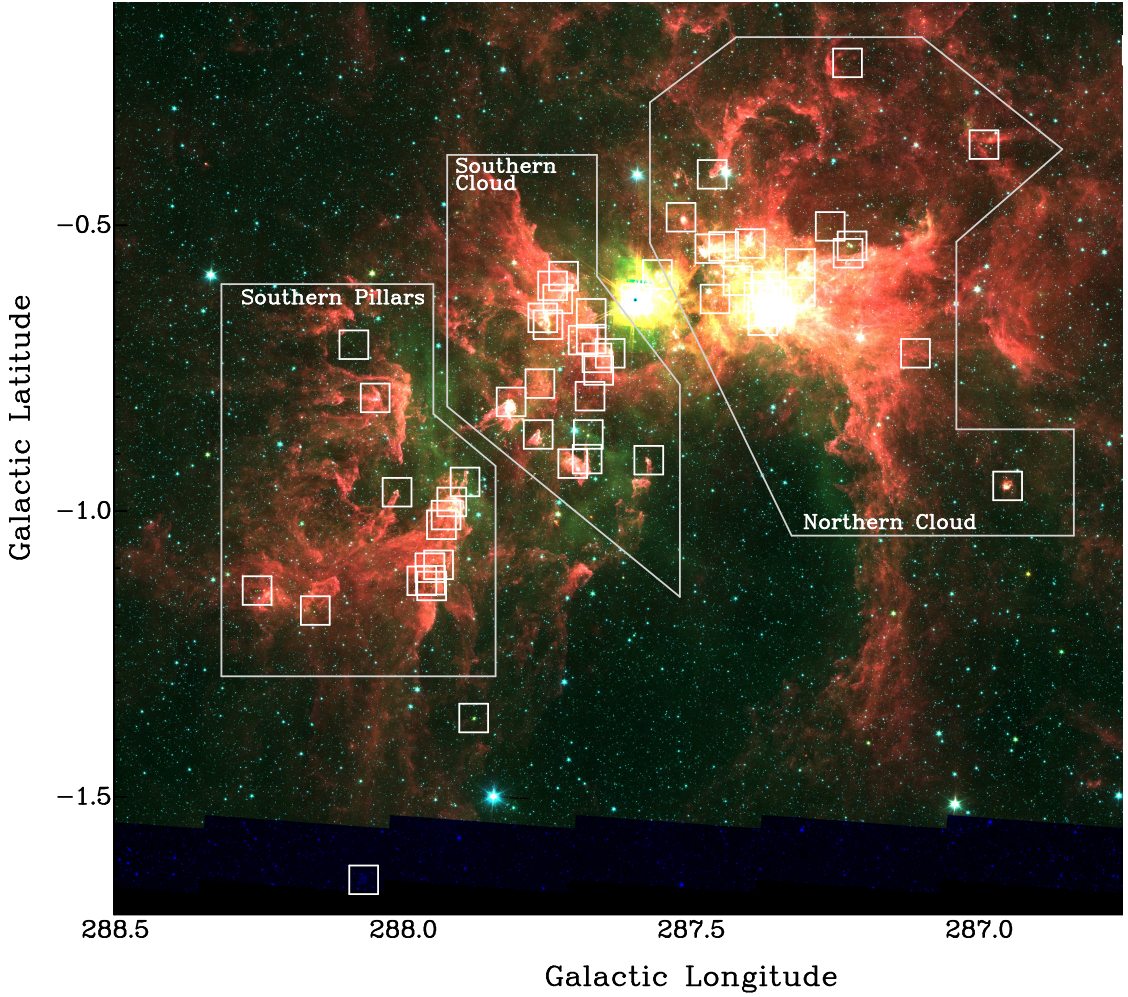}
	\caption{\textit{Spitzer}/GLIMPSE colour composite of the Carina Nebula region (Red: 8 $\mu$m, green: 4.5 $\mu$m, blue: 3.6 $\mu$m). White boxes show the extent of the 61 regions mapped with Mopra. Polygons mark the extent of the Southern Pillars, Southern Cloud and Northern Cloud regions in Carina.}
             \label{Figsources}%
 \end{figure*}

As can be seen in Fig.~\ref{Figsources}, the clumps are distributed over both the Southern and Northern Clouds as well as the Southern Pillars within the Nebula. The targets also exhibit a range of infrared properties, detected in the \textit{Spitzer} GLIMPSE bands \citep{Benjamin-2003}, which suggests they span a range of evolutionary stages. However, most of the clumps are associated with bright regions of infrared emission, often showing an 8 $\mu$m excess in the \textit{Spitzer} GLIMPSE images, which is typically associated with PDRs.

\subsection{Observing parameters}

Each of the target 61 ATLASGAL clumps was observed using the Mopra telescope and the wide-band Mopra Spectrometer (MOPS) during 17 separate observing days between 2014 Oct 16 and 2015 July 20, and between 2018 April 30 and May 2. We used the same frequency configuration as that used by the MALT90 survey \citep{jackson-2013}, allowing 16 spectral lines near 90 GHz (see Table~\ref{tab:target_lines}) to be observed simultaneously. The combination of these molecular transitions span a large range of excitation energies (from 5 to 200 K) and critical densities (from 10$^5$ to 10$^6$ cm$^{-3}$), tracing distinct physical conditions within the clumps such as high-density gas, hot core chemistry, PDRs, shocks, outflows and ionised gas.

For each of the clumps, we obtained 3\arcmin $\times$ 3\arcmin\ maps using on-the-fly mapping \citep[as in the MALT90 survey][]{jackson-2013}. In order to obtain adequate sensitivity, and to minimise striping artefacts, we mapped each source twice, scanning once in Galactic longitude and once in latitude. The telescope scanning rate was 3\farcs29 s$^{-1}$ and data was written every two seconds (or every 6\farcs58). Adjacent scanning rows were spaced by 12\arcsec\, making these observations oversampled at every frequency (the HPBW at 92 GHz is 37\arcsec). A reference observation was made at the end of each scanning row, and was offset by $-$1 degree in Galactic latitude from the target position. The system temperature was measured twice during each $\sim$30 minute map using a paddle scan, which consisted of a room-temperature paddle, that is inserted into the optical path of the telescope. The MOPS spectrometer was configured in its zoom mode which provided 16 $\times$ 137.5 MHz observing bands, each with 4096 spectral channels, corresponding to velocity channels of 0.11~\kms and a velocity coverage of $\sim$450~\kmsns.

At the start of each observing session, and between every 2 maps ($\sim$once an hour), the pointing of the telescope was corrected by observing a nearby SiO maser, RW Vel, which has a precisely known position. To ensure that the system was working and that the data were consistently calibrated from day to day, a source showing strong emission in a number of the target lines, G\,300.968+01.145, was observed each day. This test source was also observed during the MALT90 observations, ensuring that the calibration is also consistent between the two data sets.

\begin{table*}
\caption{Spectral lines targeted in these observations \citep[as in the MALT90 observations;][]{jackson-2013}.}
  \begin{tabular}{lcrrl} \hline
{\bf Transition}       &	{\bf Rest freq} & {\bf E$_u$} &{\bf  n$_{crit}$}	& {\bf Type of tracer} \\ 
      &	{\bf (MHz)} &	{\bf (K)} & {\bf (cm$^{-3}$)} \\ \hline
HCO$^+$ (1$-$0) 		&	89188.526 	& 4.28& $2 \times 10^5$ &	Density, kinematics \\
HCN (1$-$0) 			&	88631.847	& 4.25& $3 \times 10^6$ &	Density\\
HNC (1$-$0) 	 		&	90663.572 	& 4.35& $3 \times 10^5$ & 	Density, cold chemistry\\
H$^{13}$CO$^{+}$ (1-0) 	&	86754.330 	& 4.16& $2 \times 10^5$ &	Optical depth, column density\\
HN$^{13}$C (1$-$0) 		&	87090.859 	& 4.18& $3 \times 10^5$ &	Optical depth, column density\\
N$_2$H$^+$ (1$-$0)		&	93173.772	& 4.47 & $3 \times 10^5$ &	Density, chemically robust		\\
C$_2$H (1$-$0) 3/2$-$1/2&	87316.925 	& 4.19& $4 \times 10^5$ &	Cold gas, Photo dissociation region\\
SiO (2$-$1)			&	86847.010 	& 6.25& $2 \times 10^6$ &	Shocks, outflows\\
$^{13}$CS (2$-$1) 		&	92494.303 	& 6.66 & $3 \times 10^5$ &	Optical depth, column density\\
$^{13}$C$^{34}$S (2$-$1)&	90926.036 	& 7.05& $4 \times 10^5$ &	Optical depth, column density \\
CH$_3$CN 5(0)$-$4(0) 	&	91987.086 	&20.35& $4 \times 10^5$ &	Hot core       \\
HC$_3$N (10$-$9) 		&	90978.989 	&24.01& $5 \times 10^5$ &	Hot core       \\
HC$^{13}$CCN (10$-$9) 	&	90593.059 	&24.37& $1 \times 10^6$ &	Hot core\\
HNCO 4(0,4)$--$3(0,3) 	&	87925.238	&10.55& $1 \times 10^6$ &	Hot core\\
HNCO 4(1,3)$--$3(1,2) 	&	88239.027 	&53.86& $6 \times 10^6$ &	Hot core\\
H41$\alpha$			&	92034.475 	& &  &	Ionised gas    \\

\hline
\end{tabular}\label{tab:target_lines}
\end{table*}

\subsection{Data reduction}

The data were reduced using a modified version of the same automatic Python script used by MALT90, utilising the ATNF packages \texttt{asap}, \texttt{livedata}, and \texttt{gridzilla} to perform the data reduction \citep{jackson-2013}. In order to minimise artefacts due to noise spikes and ripples, the reference spectrum was smoothed with an 11-channel Hanning smooth before subtraction. For consistency with the MALT90 observations, each spectrum was gridded to a velocity resolution of 0.11 km s$^{-1}$ during the processing. 

The spectra were then co-added and gridded into a uniform 9\arcsec\ grid. Then, a 12\arcsec\ Gaussian spatial smooth was applied, resulting in an angular resolution of 38\arcsec\ for each orthogonal map. Finally, each orthogonal map was combined using the system temperature as a weight to produce a single map for each of the 16 target spectral lines. 

The resultant 3\arcmin $\times$ 3\arcmin\ maps had a typical rms noise of 0.06 to 0.1 K (1.1 to 1.9 Jy/beam at 90 GHz).

\section{Results}

\subsection{Properties of the molecular line emission}

Of the 16 spectral lines observed toward the 61 clumps in Carina, we only detected (above the 3$\sigma$ rms noise limit) molecular line emission from the HCO$^+$, N$_2$H$^+$, HCN, HNC and C$_2$H molecules. These lines have been typically the most frequently detected in other similar line surveys with the Mopra telescope \citep[e.g.][]{Sanhueza-2012}. Tentative (at the 2$\sigma$ rms noise level) detections were made of the H$^{13}$CO$^+$ molecule toward some of the clumps, but we did not detect any emission from any of the other 10 spectral lines observed with Mopra.

For each molecular line detected (above the 3$\sigma$ rms noise limit) we determined the line properties using the \texttt{pyspeckit} package \citep{Ginsburg-2011}. For molecules without hyperfine components we fitted Gaussian profiles, determining the peak antenna temperature (T$_A^*$), the velocity ($v_{LSR}$) and line velocity dispersion ($\sigma$)  (see Tables \ref{table:summary1} and  \ref{table:summary}). The full width at half maximum of the line (FWHM) is related to $\sigma$ via FWHM$=2\sqrt{2 \rm{ln} 2}~\sigma$. For sources where more than one cloud was observed along the same line of sight, we fitted one Gaussian to each line component. In these cases we have reported the Gaussian fitted line parameters for each of the velocity components.

For the HCN and N$_2$H$^+$ molecular emission that showed hyperfine structure, we simultaneously fitted each hyperfine component. For the HCN molecule we report the antenna temperature (T$_A^*$), the velocity ($v_{LSR}$) and line velocity dispersion ($\sigma$) of the main central line. For the N$_2$H$^+$ molecule we determined the opacity, the excitation temperature, the velocity ($v_{LSR}$) and line velocity dispersion ($\sigma$) from the fit. 

The line velocity dispersions have values ranging from 0.2 to 6 km $^{-1}$. These values are larger than the thermal velocity dispersion of the HCO$^+$, N$_2$H$^+$, HCN, HNC and C$_2$H molecules ($\sigma_{th}\sim0.1$ km s$^{-1}$, assuming 30 K). This suggests that the line velocity dispersion values are dominated by non thermal motions, such as turbulence within the clumps.

\begin{figure*}
   	\centering
   	\includegraphics[trim={0.5cm 1.0cm 0cm 3cm}, clip,width=0.9\textwidth]{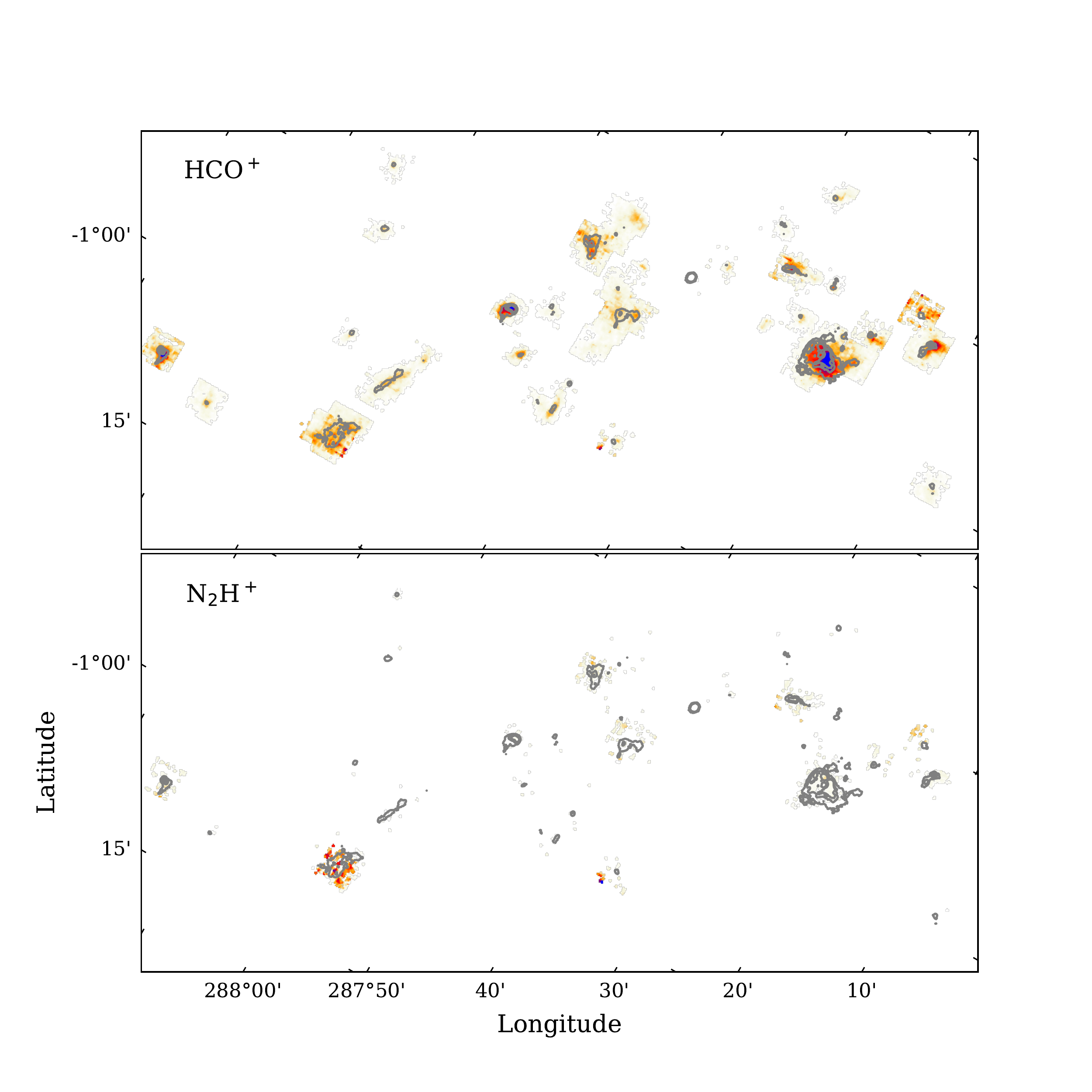}
	\caption{Overview of the HCO$^+$ and N$_2$H$^+$ molecular line emission detected towards Carina. This image shows all of the moment maps where molecular line emission was detected. Overlaid on each map are contours of the dust continuum emission observed as part of the ATLASGAL survey. Contours are 3, 6, 9 $\sigma$ of the ATLASGAL rms maps.}
             \label{Figdetection}%
 \end{figure*}

Using a 6~\kms velocity range centered on the fitted molecular line v$_{LSR}$, integrated intensity maps (moment 0) of the molecular line emission defined as $M_0=\int{ T_A^*(\nu) d\nu}$ were created. This velocity range was selected as it covers the whole line emission. To create the integrated intensity maps we excluded emission below 1$\sigma$ rms. Figure \ref{Figdetection} shows the combined integrated intensity maps for the HCO$^+$ and N$_2$H$^+$ emission for all the target regions. Appendix A shows the remaining integrated intensity maps for the detected molecules, as well as the spectra and fitted lines.

\subsection{Physical properties of the clumps}

Following the same approach used in the MALT90 survey \citep{Contreras-2017}, we used masks generated by the ATLASGAL compact source catalogues \citep{Contreras-2013a, Urquhart-2014} to determine the physical extent of each clump. We then used the column density and temperature maps computed by \citep{Rebolledo-2017} across Carina from a SED fitting of the dust continuum emission observed by \textit{Herschel}, to determine the mass and average temperature of each clump. 

The uncertainties in the column density and temperature for each clump are derived as a composite between the accumulated error for each pixel and the standard deviation estimated from all the pixels associated with the clump.   To have a more reliable estimation of the average dust temperature in the clumps, and considering that the dust temperature does not change significantly across the clumps, in the dust temperature determination we only included the pixels where the SED fitting algorithm yields $\chi^2$ values below 20, thus removing regions where our SED model is not a good description of the infrared flux spectral distribution. For more detail about the SED fitting procedure and error estimation, please refer to \citet{Rebolledo-2016}.

Table \ref{tab:target_mass} shows the physical properties for each clump. The mass of the clumps range from 14$\pm$2 to 3080$\pm$270 M$_\odot$ with a mean value of 214$\pm$5 M$_\odot$ (only one source has a a mass $>$ 1000 M$_\odot$). Using these masses, and the sizes from the ATLASGAL compact source catalogue, we derived the volume density of each clump, assuming a spherical shape. Their number densities range between (8$\pm$0.7)$\times$10$^3$ and $(1.1\pm0.1)\times10^6$ cm$^{-3}$, with a mean value of (1.6$\pm$0.5)$\times10^5$ cm$^{-3}$.

Assuming the clumps have a star formation efficiency of 30\% and following the empirical relationships from \citep{Larson-2003} we can estimate the mass of the most massive star that can be formed in the clump. According to this estimation, only 10 of the 61 clumps ( 16\%) will eventually host a high-mass star (star with more than 8 M$_\odot$). This percentage is small considering that in the rest of the Galaxy, following the same assumptions, 82\% of the clumps observed by ATLASGAL will potentially form high-mass stars \citep{Contreras-2013a}. Thus, although most of the clumps are dense enough to form stars, not many have the conditions to form high-mass star in the future. 

The temperatures of the clumps are given in Table \ref{tab:target_mass}, and range from 20.3$\pm$0.4 to 42.3$\pm$1.8 K with a mean temperature of 26.6$\pm$0.1 K. The hottest clump corresponds to the position of $\eta$ Carina, thus, without considering this region, the rest of the clumps have temperatures up to 34.5 K. In general, for a given clump its temperature ranges within one degree from its reported mean temperature. Comparing these temperatures with the typical temperatures observed for the total MALT90 sample, we find that in general these clumps are relatively warm and they have similar temperatures to those found for HII (mean value of 23.7 K, standard deviation of 5.2 K) and PDRs (mean value of 28.1 K, standard deviation of 5.9 K) regions in the rest of the Galaxy \citep{Guzman-2015}.

\subsection{The dense gas in Carina}

We compared the distribution of the dense gas, traced by the molecular line emission integrated intensity, to the dust column density and the temperature of the dust computed by \citep{Rebolledo-2017}. The integrated intensity maps of the HCO$^+$ emission shows that the dense gas is correlated with dust continuum emission from ATLASGAL (as shown in Fig.~\ref{Figdetection}). We do not detect much N$_2$H$^+$ emission, suggesting that there is no much cold dense gas in this region, or the regions containing N$_2$H$^+$ are very compact, thus we are not able to detect them with Mopra due to beam dilution effects. 

\citet{Rebolledo-2016} determined the column density toward Carina using three different methods:  1). From the $^{12}$CO molecular line emission assuming a constant $X_\mathrm{co}$ factor across Carina. 2). From $^{13}$CO emission using an LTE approximation and assuming an abundance ratio, and 3) from a SED fitting of the dust continuum emission observed by \textit{Herschel}, assuming a constant Gas-to-Dust ratio of 100. \citet{Rebolledo-2016} found that by comparing the material traced by the CO emission to the total mass reservoir of Carina, the fraction of the molecular gas changes across the Carina complex.  In the case of the Northern Cloud, the molecular gas corresponds to $\sim$ (80$\pm40$)\% of the total mass in the region.  On the other hand, the Southern Cloud whose projected distance to the massive star clusters is similar to the Northern Cloud only has a molecular gas fraction of $\sim$ (50$\pm25$) \%.  Finally, the gas located in the Southern Pillars is $\sim$ (70$\pm$35)\% molecular, which is expected considering that the Southern Pillars are located farther away from the massive stars.  

\begin{figure*}
   	\centering
   	\includegraphics[trim={0cm 0.1cm 0cm 1.3cm}, clip,width=0.45\textwidth]{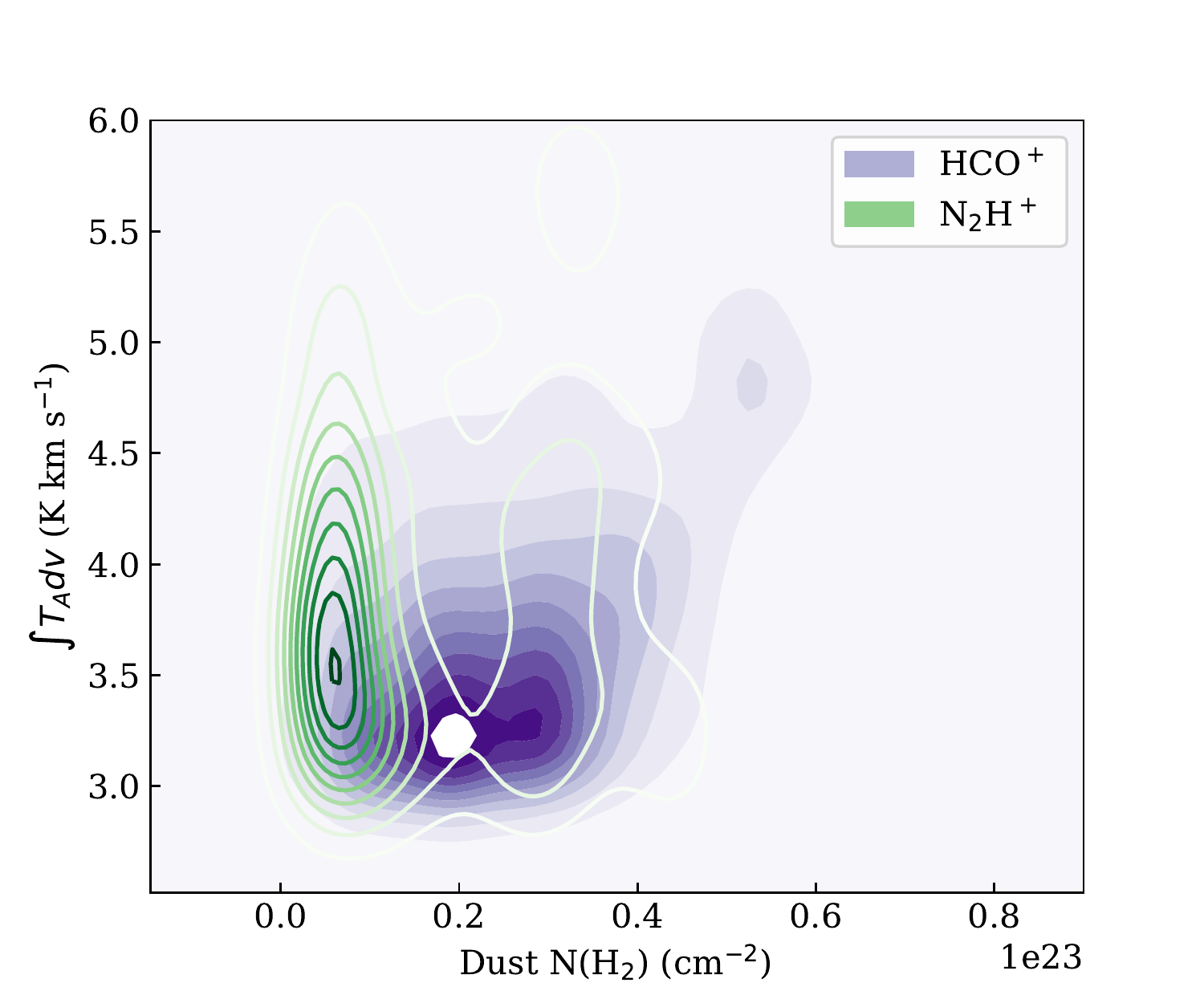}
    \includegraphics[trim={0cm 0.1cm 0cm 1.3cm}, clip,width=0.45\textwidth]{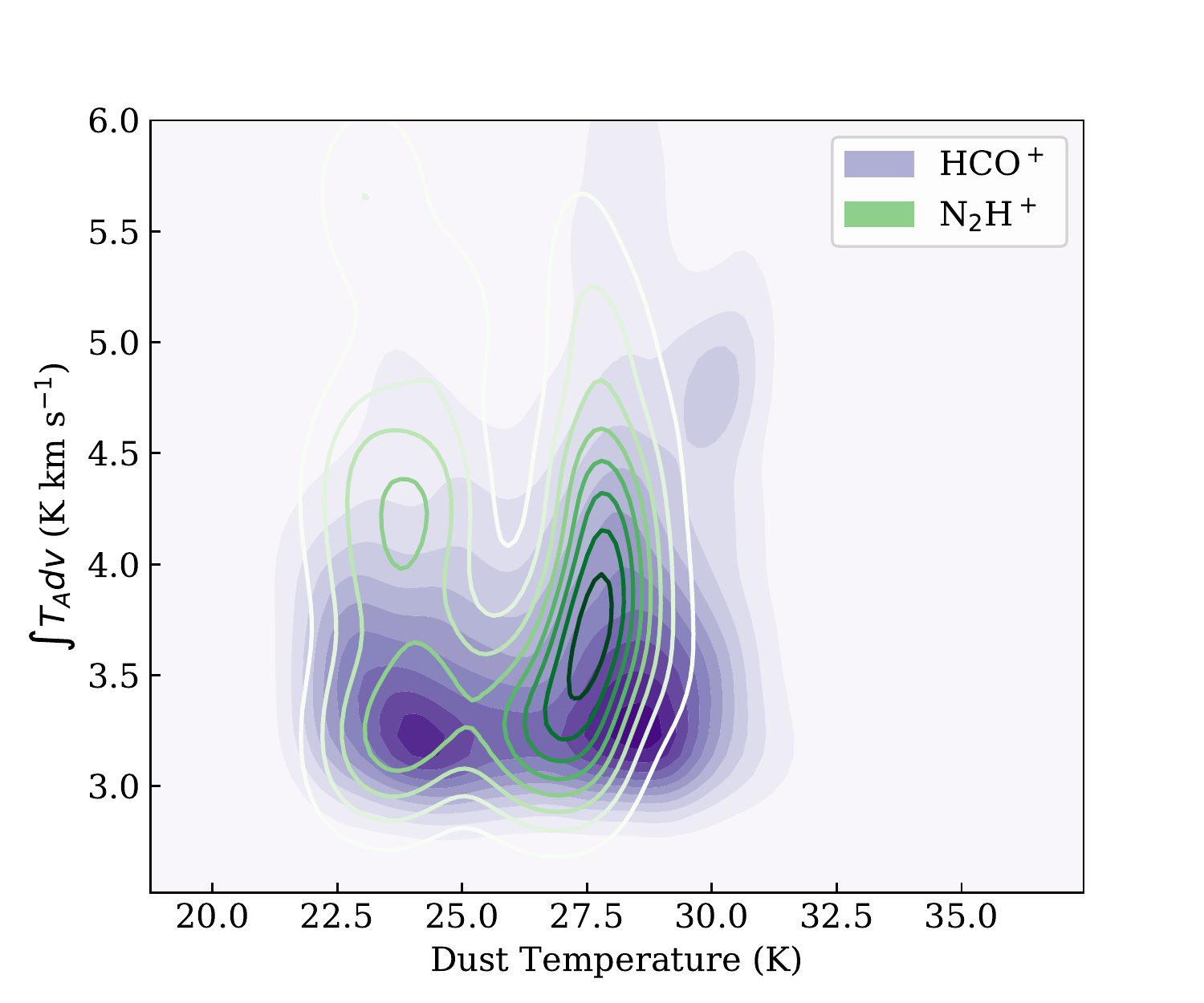}
    \includegraphics[trim={0cm 0.1cm 0cm 1.3cm}, clip,width=0.45\textwidth]{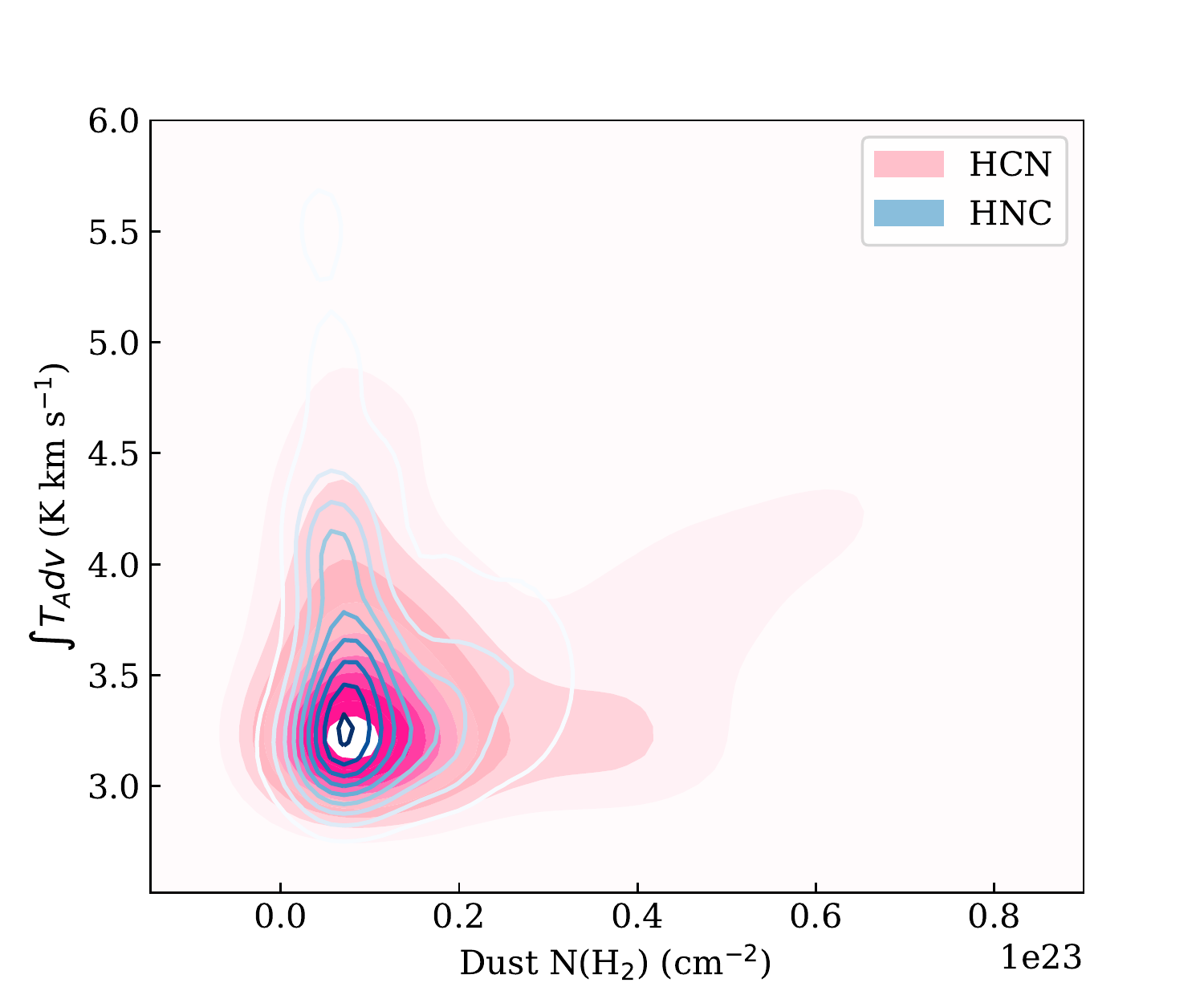}
    \includegraphics[trim={0cm 0.1cm 0cm 1.3cm}, clip,width=0.45\textwidth]{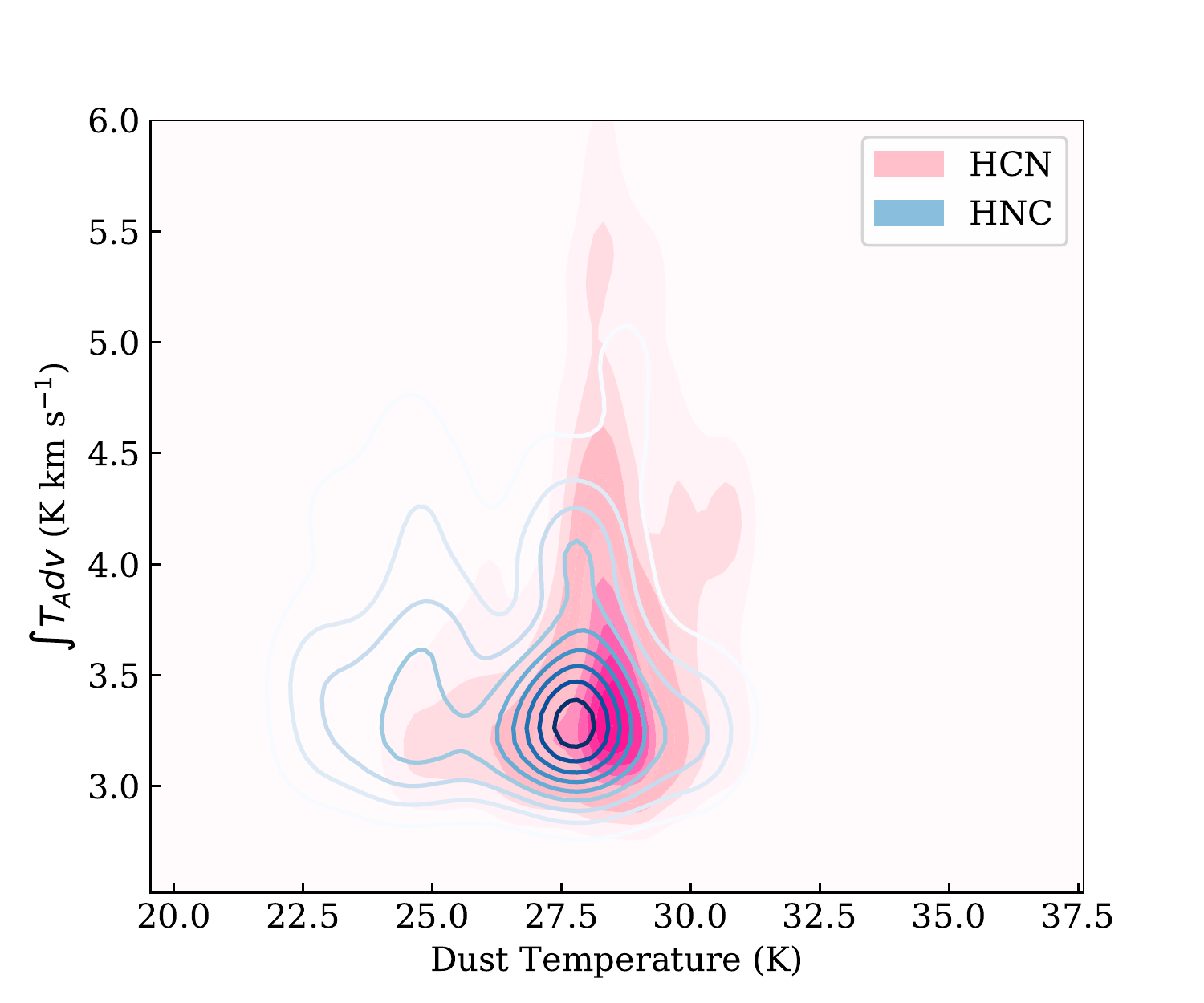}
	\caption{Density plots of the integrated intensity versus the dust column density (left panels) and temperatures (right panels). Column density and temperatures correspond to a fit to the dust continuum emission observed by \textit{Herschel} \citep{Rebolledo-2017}. Top panels show the integrated intesity for HCO$^+$ and N$_2$H$^+$, where it can be seen that N$_2$H$^+$ is detected in regions with slightly lower dust column density compared to the regions where HCO$^+$ is detected. Lower panels show the HCN and HNC emission. Both molecules are detected toward regions with similar dust column density values, however HCN is detected toward regions where the dust temperature is slightly higher than the regions where HNC is detected.}
             \label{Fig:densegas}%
 \end{figure*}

In the regions observed with Mopra, 90\% of the gas has column densities $<8\times10^{22}$ cm$^{-2}$. The most commonly detected molecular emission, HCO$^+$ (1-0), seems to trace slightly denser and hotter material than the other molecules (see Fig. \ref{Fig:densegas}). HCO$^+$ trace 85\% of the gas with more than $>2\times10^{22}$ cm$^{-2}$, and all the gas with column densities $>5\times10^{22}$ cm$^{-2}$. HCN, HNC and N$_2$H$^+$ typically trace the regions where the column density is $\sim10^{22}$ cm$^{-2}$, and temperatures of $\sim 25 - 28$ K. In general, we found that the molecular line emission traces well the higher column density gas in Carina. 

\subsection{Molecular emission of the Carina clumps in the Galactic context}

We compared the detection rates of the clumps in Carina to detection rates of the 3246 clumps observed in the Galactic plane as part of the MALT90 survey \citep{Rathborne-2016,jackson-2013,Foster-2011}. 

We detect HCO$^+$ emission toward 93\% of the Carina clumps, which is comparable to the HCO$^+$ detection rate in MALT90 of 88.5\%. The 32\% detection rate of the C$_2$H molecule in Carina is also similar to the MALT90 detection rate of 37\%. The detection rates of the three other lines that we detected in Carina, HNC, HCN and N$_2$H$^+$, showed much lower detection rates at 52\%, 54\%, and 23.0\%, compared to the detection rates in the MALT90 sample of 87.\%, 77.6\% and 76.5\%.

These detection rates are very different to the detection rates of the total MALT90 sample, which contains clumps that are in a range of evolutionary stages from pre-stellar, protostellar and more evolved clumps located in HII Regions and PDRs. This is surprising, considering that Carina is closer than most of the MALT90 clumps \citep[75\% of the MALT90 clumps have kinematic distances $>3.5$ kpc,][]{Whitaker-2017}

\begin{figure}
   	\centering
   	\includegraphics[trim={1cm 1.5cm 1cm 1cm}, clip,width=0.5\textwidth]{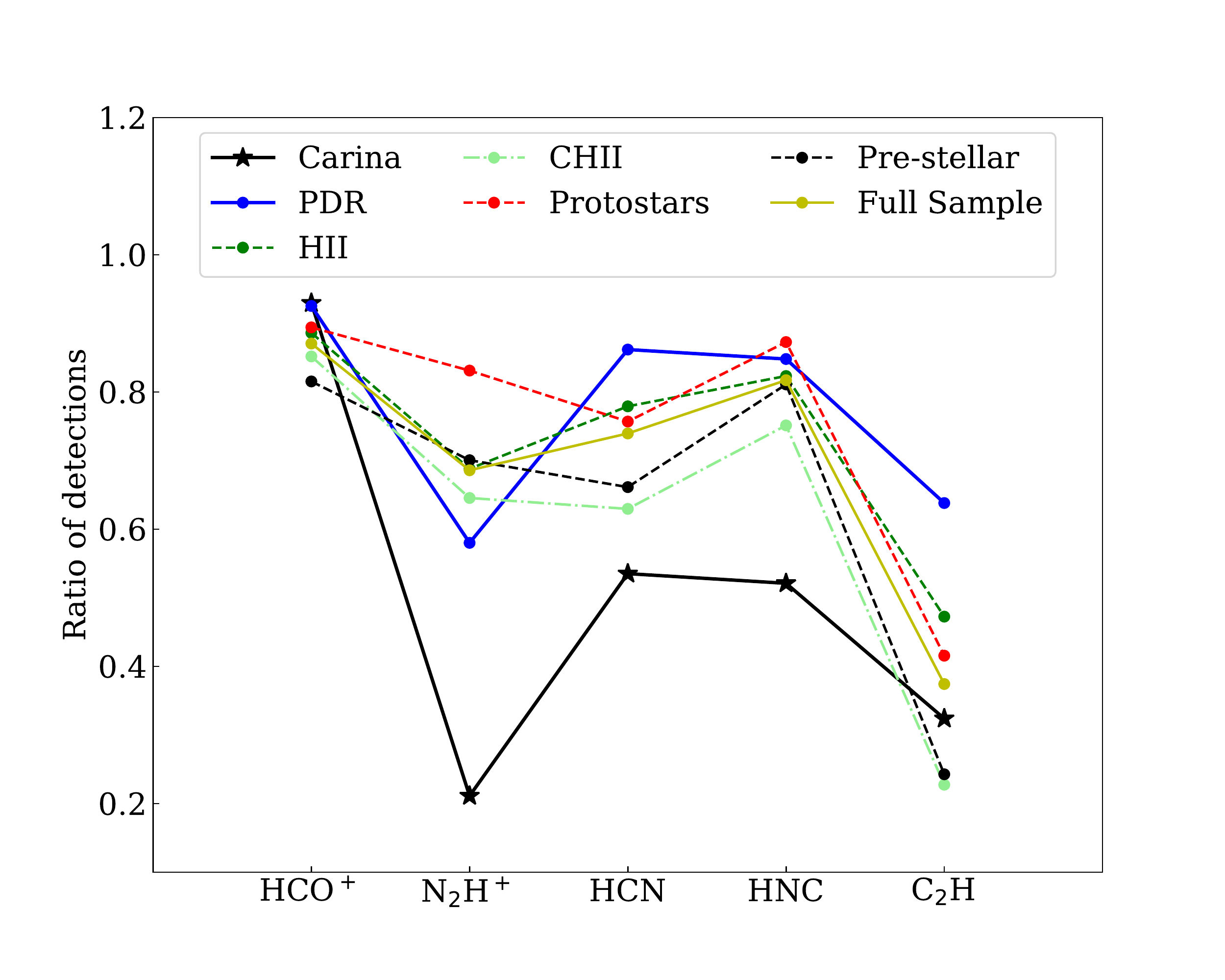}
	\caption{Detection rate of the five molecules detected toward Carina compared to the detection rates of the MALT90 survey. The clumps in Carina clearly shows the same detection pattern as the MALT90 associated with PDRs, although the detection rates for all the molecules, except HCO$^+$, are lower than the ones found for PDRs.}
             \label{Fig:detectionrate}%
 \end{figure}

To determine if the detection rates are correlated with the stage of evolution of the clumps, we compared the detection rates found for the clumps in Carina, to the MALT90 detection rates for clumps classified in the different evolutionary stages. Figure \ref{Fig:detectionrate} shows the MALT90  detection rates as function of evolutionary stage, the detection rate for the full sample and the detection rate in Carina. While the detection rates in Carina are very different in general, having a low detection rate of N$_2$H$^+$, HNC and HCN compared to the rest of the MALT90 clumps, the overall trend of the detection rates is similar to that found for the MALT90 clumps associated with PDR regions. 

This suggests that the Carina clumps follow a chemical evolution similar to the MALT90 PDR regions across the Galaxy. This is consistent with previous works that has associated the 8$\mu$m emission excess, H$_2$ and polycyclic aromatic hydrocarbon (PAH) emission observed in Carina to be consistent with PDRs \citep{Rathborne-2004,Rathborne-2002,Brooks-2000}.

\subsection{Kinematics of the clumps}

The line profiles can be used to study the kinematics within the clumps.  Self-absorption in line profiles of high critical densities of molecular tracers, such as HCO$^+$, can be explained by a warmer excitation temperature in the inner region of clumps. At the centre, the density can be higher and the HCO$^+$ emission can be thermalised; while in the outer envelope the gas density is lower, and the emission can be sub-thermal. Asymmetric self-absorbed profiles are typically associated with radial movement of the gas. In particular, blue-red asymmetries with the blue-shifted peak brighter than the red-shifted peak, are typically explained as a sign of infall, assuming the inner gas in the core has a warmer excitation temperature than the outer envelope.

Thus, to determine whether we detect any signatures of infall toward the clumps we analysed the HCO$^+$ (1-0) line profile. We found that two thirds of the the HCO$^+$ line profiles are Gaussian, suggesting that there are no radial motions within them. For 20 of the 61 clumps we detect asymmetries in their HCO$^+$ profiles (see Figure A2 in Appendix A of the online material). Of these 20 clumps, four of them only have emission detected in HCO$^+$, thus, we cannot confirm whether the asymmetry seen is due to radial motions in the clump, or due to two clumps along the same line of sight with slightly different $v_{LSR}$. In seven clumps, other detected molecules, such as HNC, do not show any self-absorbed profiles, which may be due the higher critical densities of these molecules that prevent the appearance of such profiles. For the remaining nine clumps, self-absorption is also seen in other tracers. 

These results suggest that there might be some infall in the clumps, which could be due to current star formation activity, causing the clumps to collapse gravitationally. However, since we do not detect any optically thin tracers toward these clumps we cannot confirm whether the profiles are truly due to infall or are the superposition of different clouds along the line of sight.

\section{Discussion}

\subsection{Are the ATLASGAL clumps in Carina Nebula typical?}

 \begin{figure*}
   	\centering
   	\includegraphics[trim={0cm 0cm 0cm 1cm}, clip,width=0.45\textwidth]{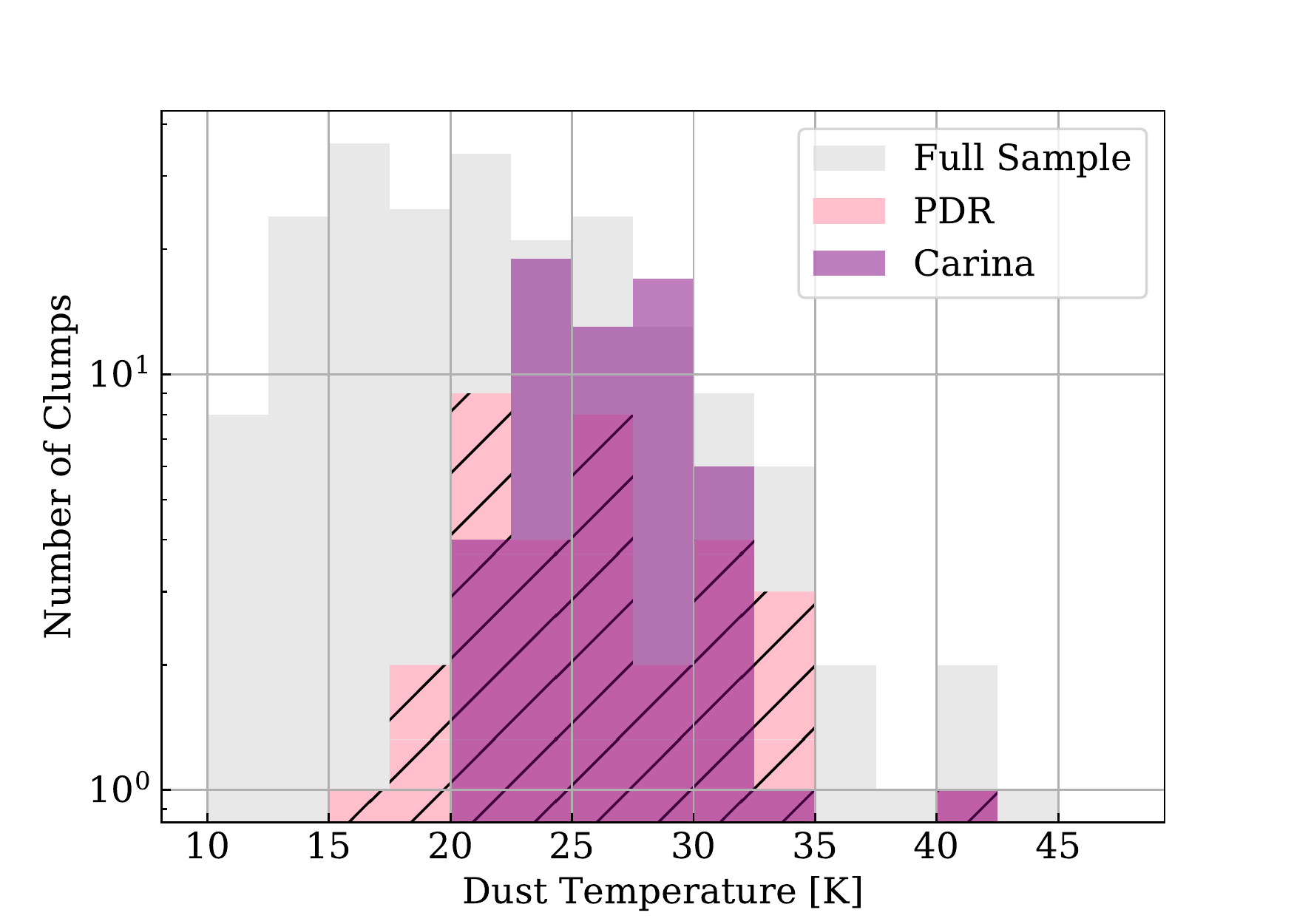}
    	\includegraphics[trim={0cm 0cm 0cm 1cm}, clip,width=0.45\textwidth]{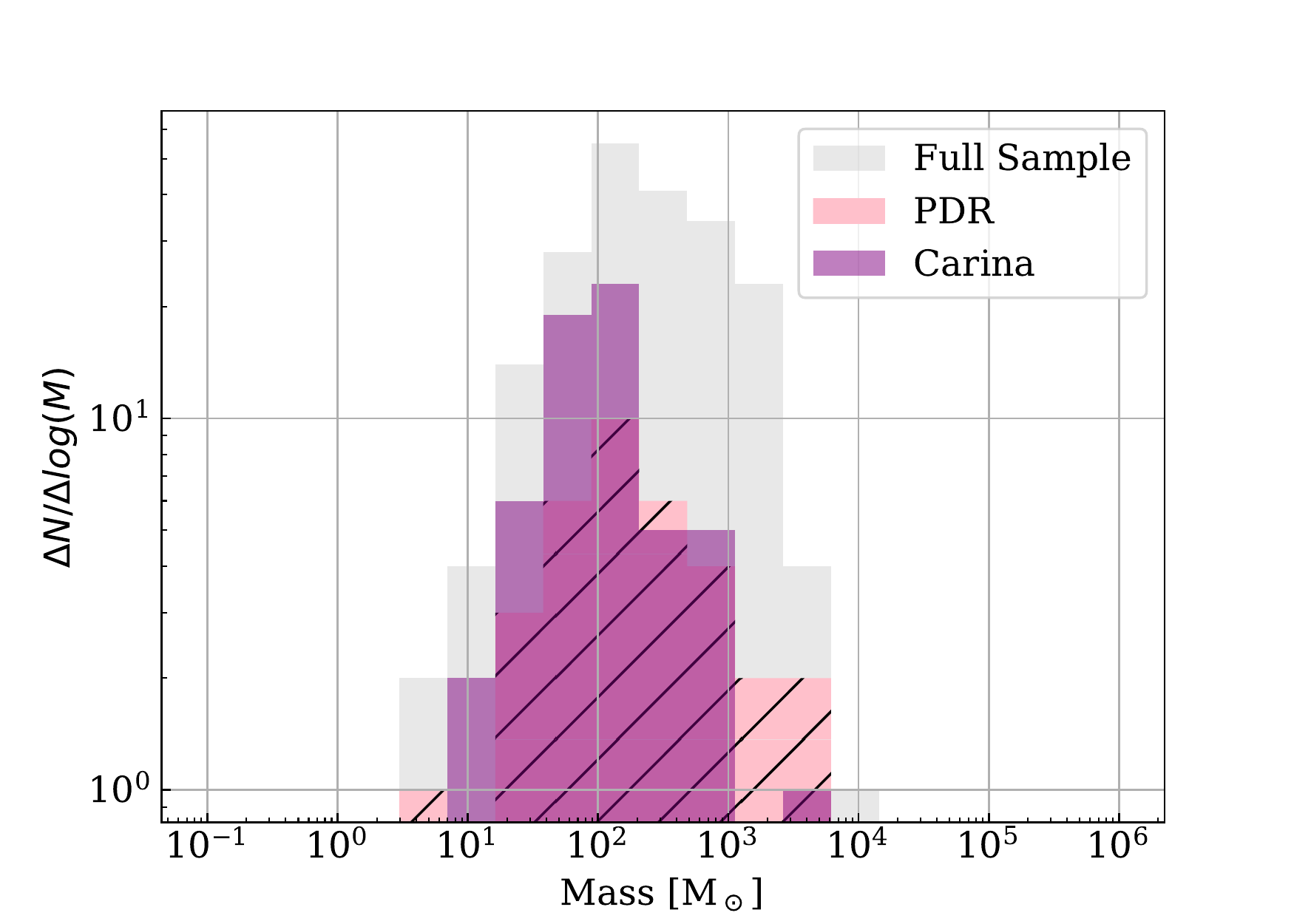}
	\includegraphics[trim={0cm 0cm 0cm 1cm}, clip,width=0.45\textwidth]{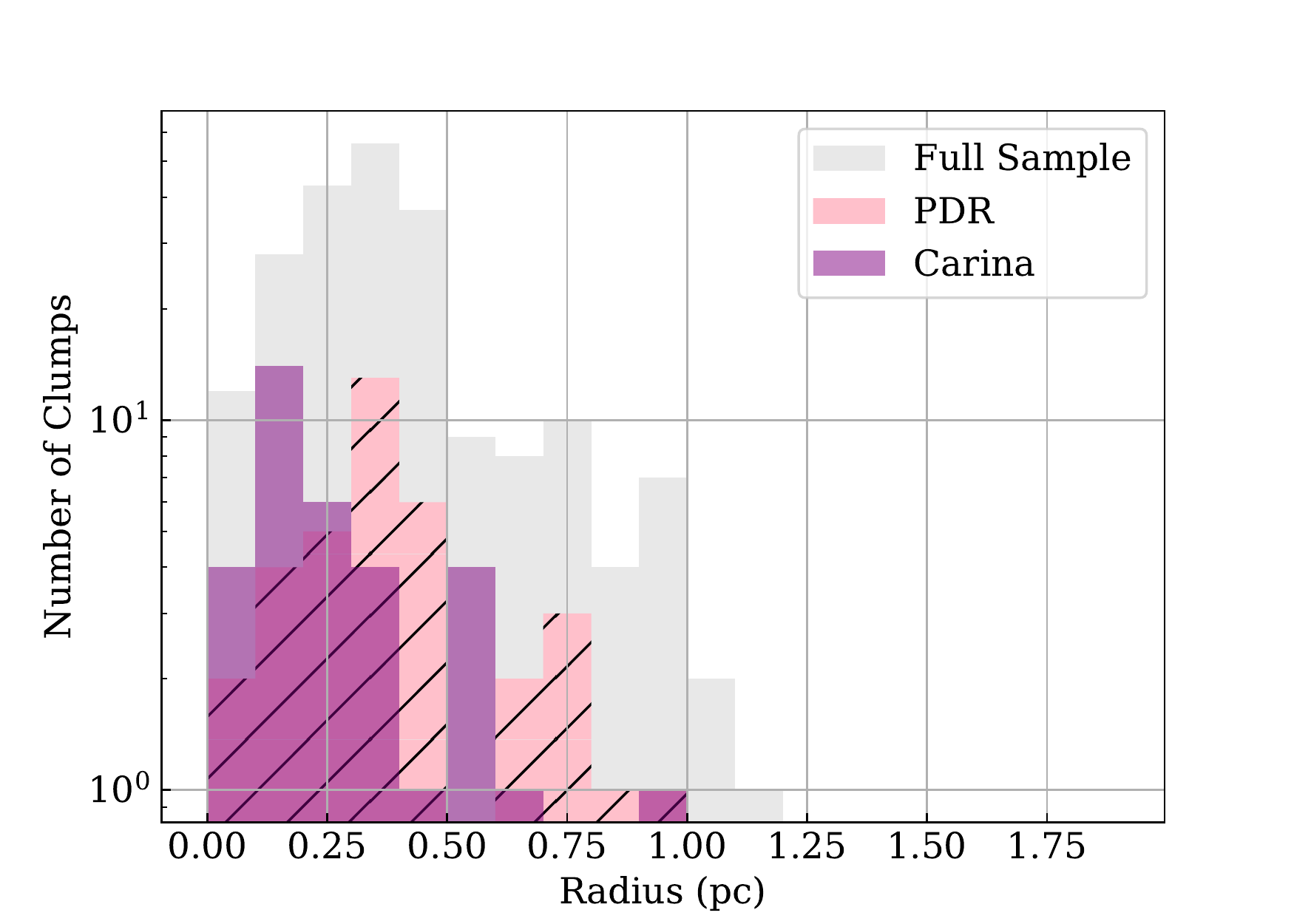}
    	\includegraphics[trim={0cm 0cm 0cm 1cm}, clip,width=0.45\textwidth]{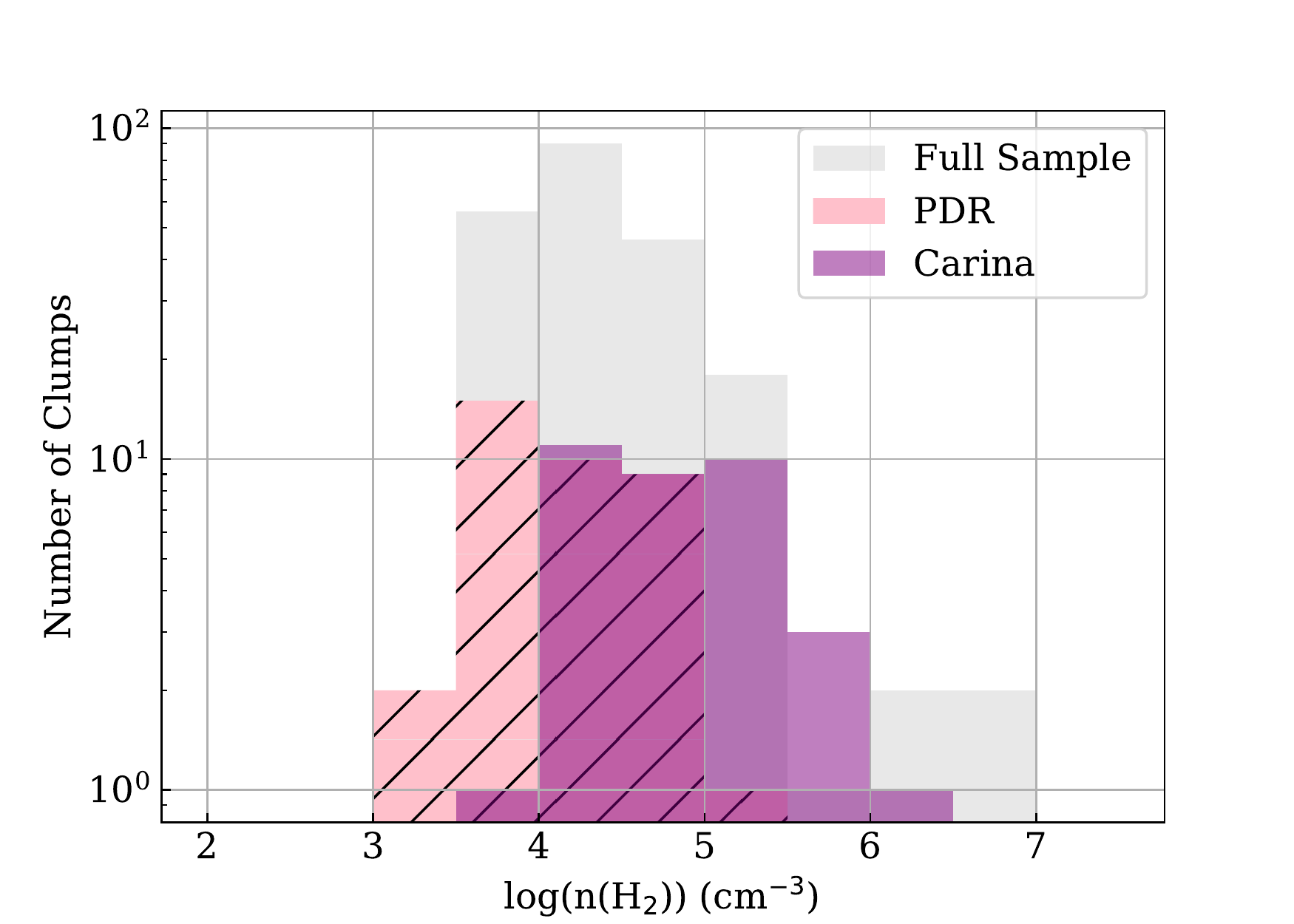}

	\caption{Histograms of the mass, temperature, radius, and volume density distribution of the clumps observed towards Carina (purple), compared with the clumps observed by MALT90 in all evolutionary stages (grey), and those associated with PDR regions (pink, dashed), both limited to those sources located within 1 and 3 kpc from us.}
    \label{Fig_masstemperature}%
 \end{figure*}

We compare the dust temperature, mass, size and volume density of the clumps in Carina with the clumps observed in the Galactic Plane by MALT90. In the following analysis, we have selected the MALT90 clumps that have a kinematic distances between 1 and 3 kpc, to ensure that the physical resolution is similar. In this manner the physical scales of the clumps, and the regions traced by the molecular line emission is similar to what is observed towards Carina. Thus, the total number of clumps used for the following analysis is 204, and they are located at an average distance of 2.48 kpc from us.

Since the detection rate of the molecular lines for the clumps in Carina follows the pattern exhibited by the clumps associated with PDR regions, we also compare the clumps in Carina with the PDRs regions located between 1 and 3 kpc of distance. This corresponds to a sample of 34 MALT90 clumps.

Figure \ref{Fig_masstemperature} shows an histogram of the physical properties found for the nearby MALT90 clumps, the subsample of nearby clumps classified as PDRs, as well as the clumps in Carina. In general, the clumps in Carina have higher temperatures than the rest of the clumps in the Galactic plane, with a mean temperature of 26.6$\pm$0.1 K, compared to the mean value of 21.2$\pm$0.4 K of the MALT90 clumps. However, they span a similar range in temperatures compared to the clumps associated with PDRs, which are the clumps with the higher mean temperature (25.7 K) in the MALT90 sample \citep{Guzman-2015}. 

Also, the clumps in Carina have lower masses (mean mass 214$\pm$5 M$_\odot$) than both the full sample of nearby MALT90 clumps (mean 508$\pm30$ mass M$_\odot$) and those nearby MALT90 sources associated with PDR regions (mean mass 554$\pm70$ M$_\odot$).  However, since the clumps are also on average smaller, with a mean radius of 0.27 pc, compared to the mean of 0.4 pc found for the MALT90 clumps, the clumps in Carina tend to have higher volume densities than the nearby MALT90 PDR regions, having a mean density of 1.6$\times10^5$ cm$^{-3}$ compared to the mean density of 2.4$\times10^4$ cm$^{-3}$.

\subsection{Chemistry anomalies in the Carina Nebula}

\begin{figure*}
   	\centering
   	\includegraphics[trim={1.8cm 0.7cm 1.5cm 1cm}, clip,width=0.59\textwidth]{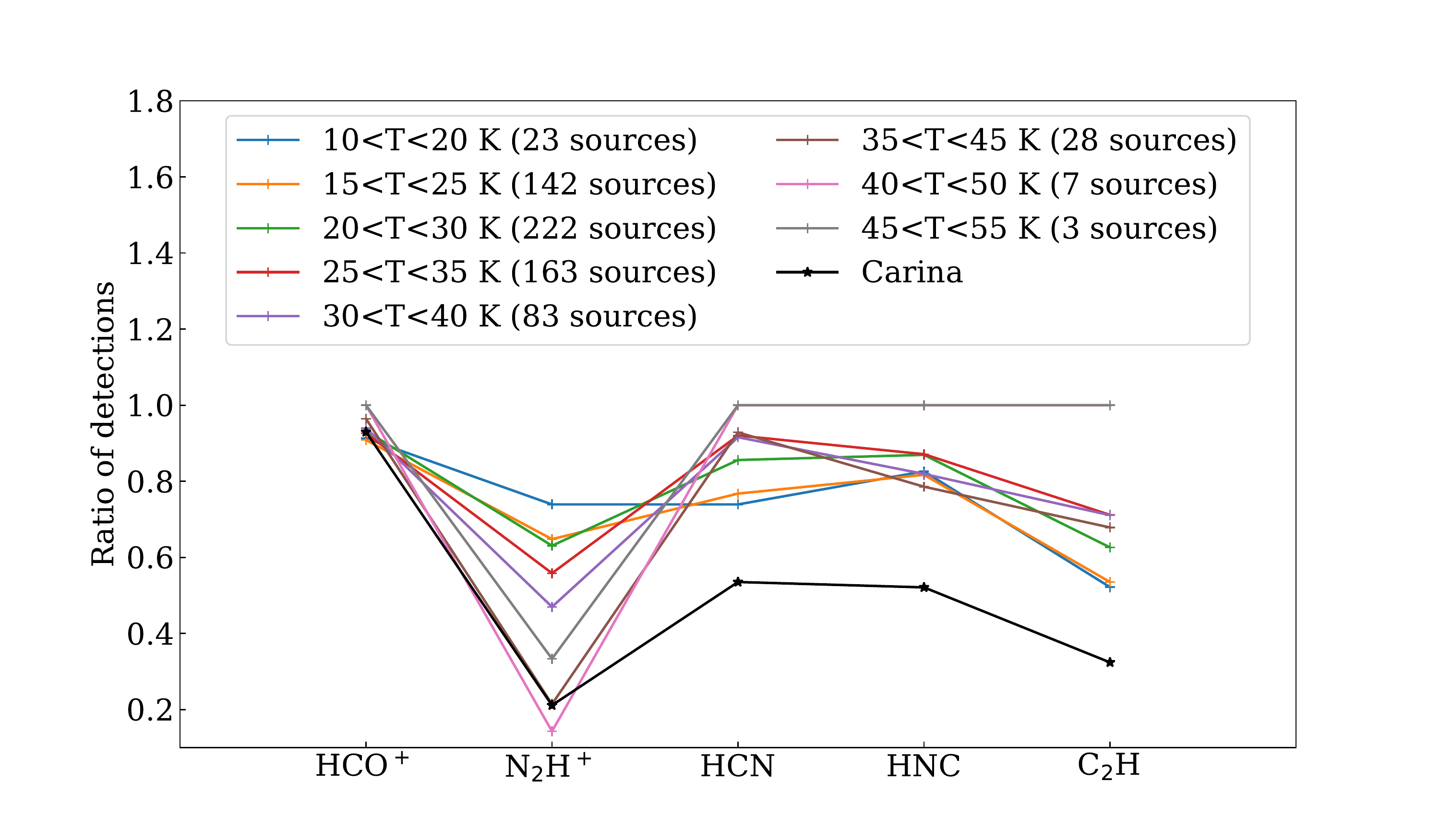}
    	\includegraphics[trim={0.2cm 0.cm 1cm 1cm}, clip,width=0.4\textwidth]{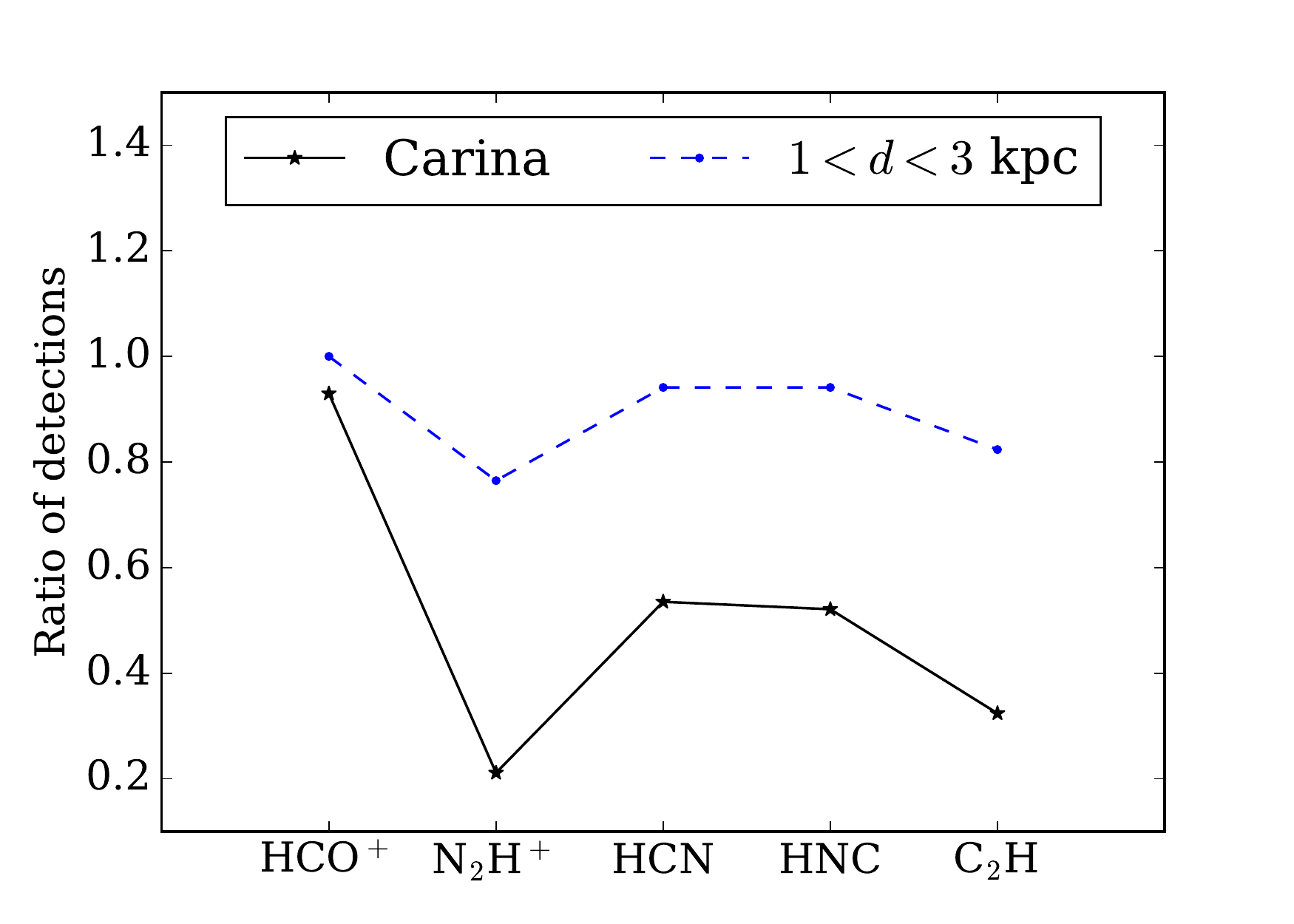}
	\caption{Detection rate of the five molecules detected toward Carina compared to the detection rates of the clumps associated to PDRs as function of the temperature (\textit{right panel}) and kinematic distance of the MALT90 clumps (\textit{left panel}).}
             \label{Fig:detectionrate2}%
 \end{figure*}

The clumps in Carina shows low detection rates for most of the molecules observed compared to the Galactic clumps observed by MALT90. Since Carina already has high-mass stars that have increased the temperature of their environment, it is possible that this has affected the detection rates observed. Since only 34 clumps are associated with PDRs between 1 and 3 kpc, to analyze the effect of the temperature we analysed the total MALT90 sample of clumps associated with PDRs.  In Figure \ref{Fig:detectionrate2} we compare the detection rates of Carina with those of the total MALT90 sample of PDR sources as function of temperature. The detection rate of N$_2$H$^+$ (1-0) does decrease with increasing the temperature of the PDRs, however the detection rates for HCN, HNC and C$_2$H are still lower in Carina.

We also compared the detection rates towards PDRs that are located within 1 and 3 kpc (Fig.~\ref{Fig:detectionrate2}) from us, to determine if by resolving some of the clumps emission we alter the detection rates of the molecular line emission. We found that the detection rates are not affected by the distance of the clumps. In fact, since Carina is closer, we would expect to detect more easily the molecular line emission. These results suggests that there might be differences in the chemistry of the clumps due to the extreme radiation observed in this region. 

\begin{figure*}
   	\centering
   	\includegraphics[trim={0cm 0cm 0cm 1cm}, clip,width=0.75\textwidth]{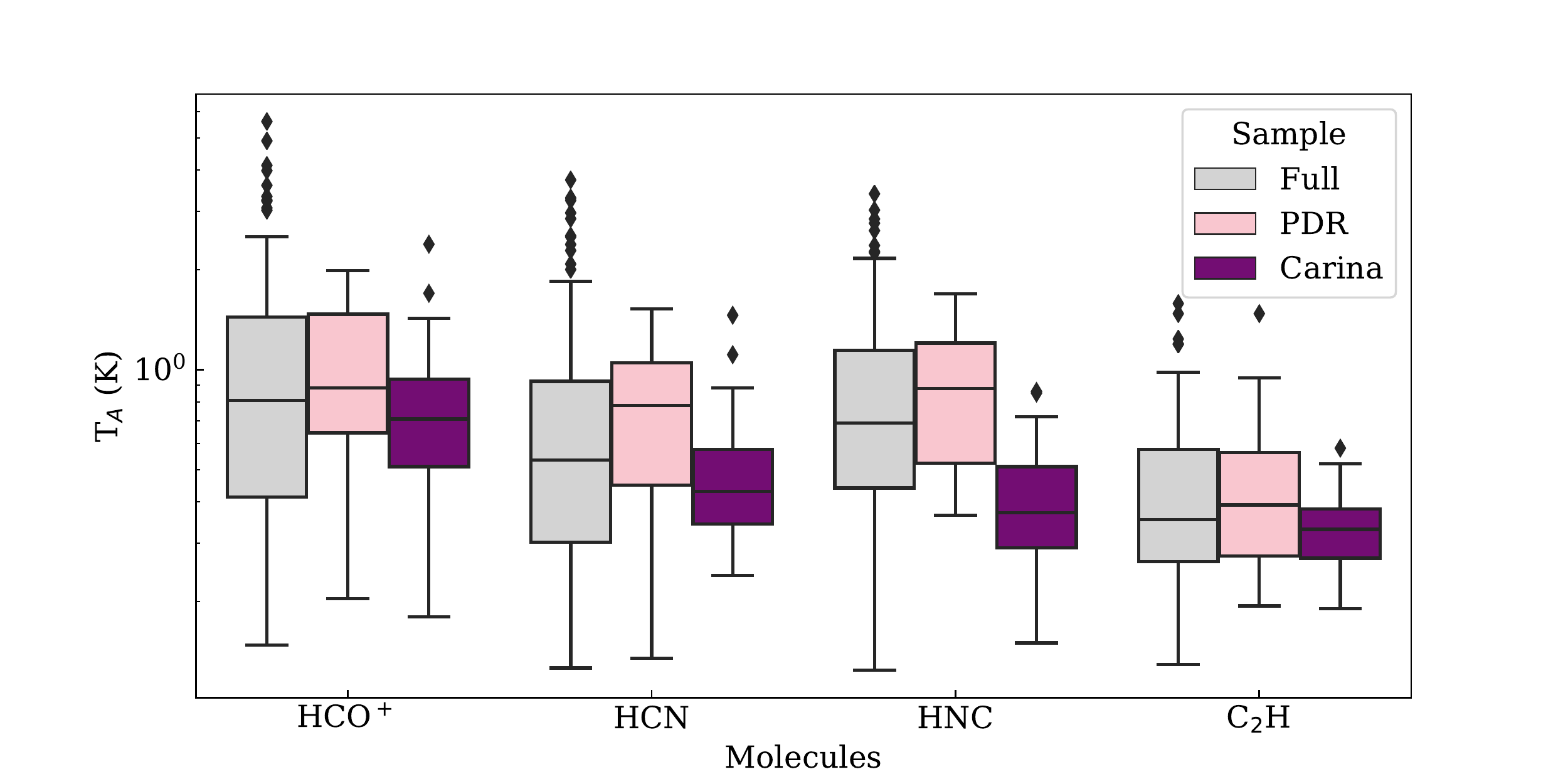}
    	\includegraphics[trim={0cm 0cm 0cm 1cm}, clip,width=0.75\textwidth]{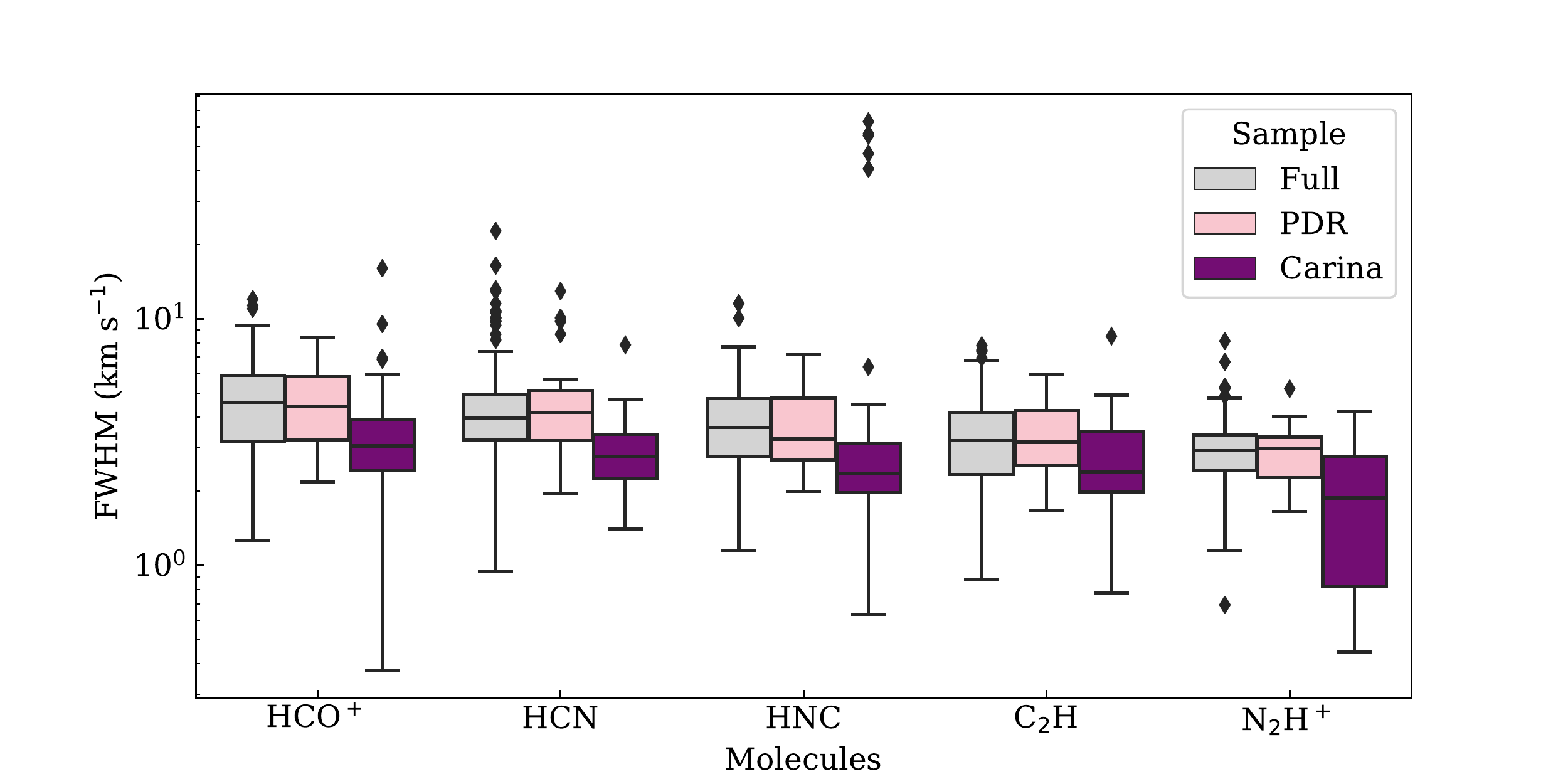}
    	\caption{\textit{Upper panel:} Peak antenna temperatures for the molecules detected in Carina (purple boxes) compared to the values obtained by the MALT90 survey for the clumps located within 1 and 3 kpc in a range of evolutionary stages (grey boxes) and those associated with PDR regions (pink boxes). \textit{Lower panel:} Line-widths of the emission, color of boxes as in the upper panel. In both panels the size of the boxes represent the second quartile of the data, the horizontal line shows the median of the values and the whiskers shows the 95\% of the data, with outliers shown as dots at either side of the boxes.}
             \label{Fig_molecules}%
 \end{figure*}

Fig.~\ref{Fig_molecules} presents a comparison of the peak antenna temperature and the line-widths of the molecules detected in Carina to both the nearby MALT90 clumps and those in the PDR category, showing that in general the peak antenna temperatures of the Carina sources are within the range of the MALT90 clumps.  However all molecules have lower mean values of peak antenna temperature than both the sample of nearby MALT90 clumps and those associated with MALT90 PDRs at similar distance to Carina. 

The line-widths of Carina detections are narrower than those found for the nearby MALT90 clumps and the PDRs. Of these molecular lines HCN is expected to have narrower line-widths than the ones found by MALT90 because, unlike MALT90, we could fit well the hyperfine structure for most of the sources. For C$_2$H and N$_2$H$^+$, although still narrower, the line-widths are more similar to that found for all the nearby MALT90 clumps and PDRs. Given that the typical line widths observed are dominated by non-thermal turbulence, for both the clumps in Carina and in the rest of the Galaxy, the narrower line-widths suggest that the clump gas is less turbulent, which might be due to their higher density compared to the MALT90 clumps. 

\begin{figure*}
   	\centering
   	\includegraphics[trim={0cm 0cm 1cm 1cm}, clip,width=0.9\textwidth]{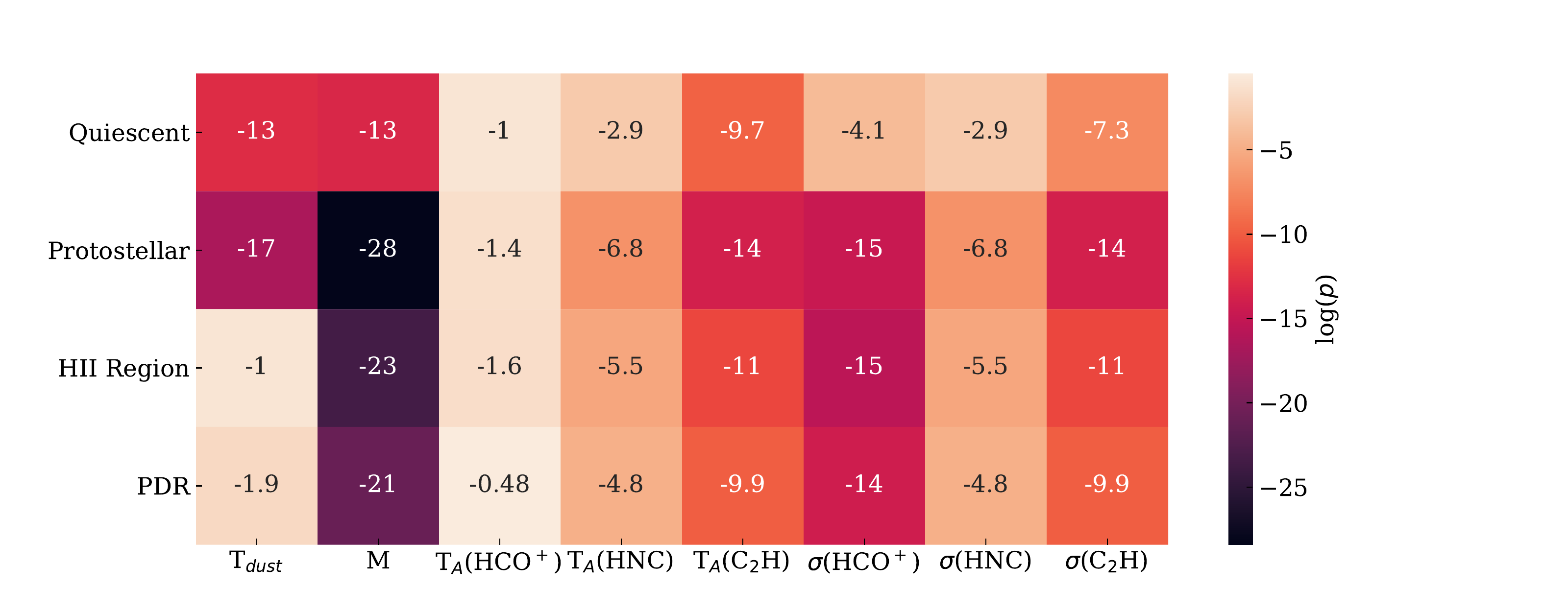}
	\caption{Heat-map of the KS test $p$ values. Each cell shows the value of the log($p$) from the KS test between the values found for dust temperature (T$_{dust}$), mass (M), molecular line peak temperature (T$_A$), and velocity dispersion ($\sigma$) between the clumps in Carina and the MALT90 clumps in different evolutionary stage classifications. If the $p$ value is small, the KS test implies that the two groups were sampled from populations with different distributions (e.g. a $p$ value of 0.01 suggest that there is 1\% probability that both samples are drawn from the same distribution). In this case, the KS test shows that all the parameters are very different between the clumps in Carina and MALT90 clumps in different evolutionary stages. The only cases where the parameters are less dissimilar are the dust temperature of clumps associated with HII Regions and PDRs, and the HCO$^+$ peak temperatures of clumps at all the evolutionary stages.}\label{kstest}%
 \end{figure*}

For completeness we also perform KS tests between the physical and chemical properties of the clumps in Carina with the nearby MALT90 clumps as function of evolutionary stage.  The KS test indicates the probability of two datasets of being sampled from populations with the same distributions. A small $p$ value (e.g. $<0.01$), indicates that we can reject the hypothesis that both samples are from the same population. Figure \ref{kstest} show the results of the $p$ values of the KS tests. We find that the properties of the clumps in Carina are from a different distribution compared to each of the MALT90 evolutionary classifications.

In general, our observations suggest that the clumps in Carina are chemically and physically different to those found in much of the Galaxy.

\subsection{Effect of high-mass stars on its environment}

The differences in the physical and chemical properties observed in Carina may be due to the proximity of the clumps to the high-mass stars present in this region. 

\begin{figure*}
   	\centering
   	\includegraphics[trim={0.5cm 0.cm 0cm 1cm}, clip,width=0.9\textwidth]{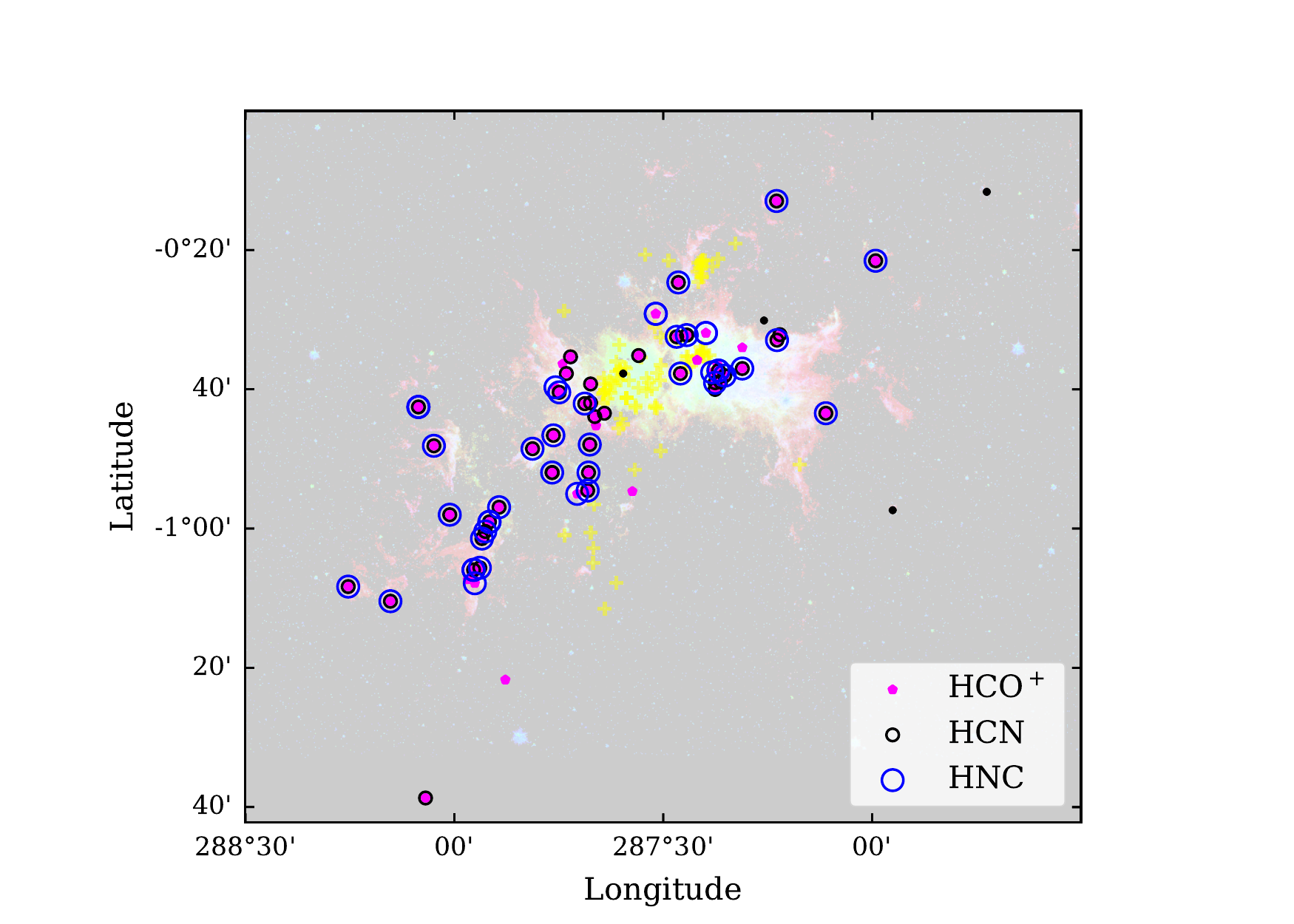}
	\includegraphics[trim={0.5cm 0.cm 0cm 1cm}, clip,width=0.9\textwidth]{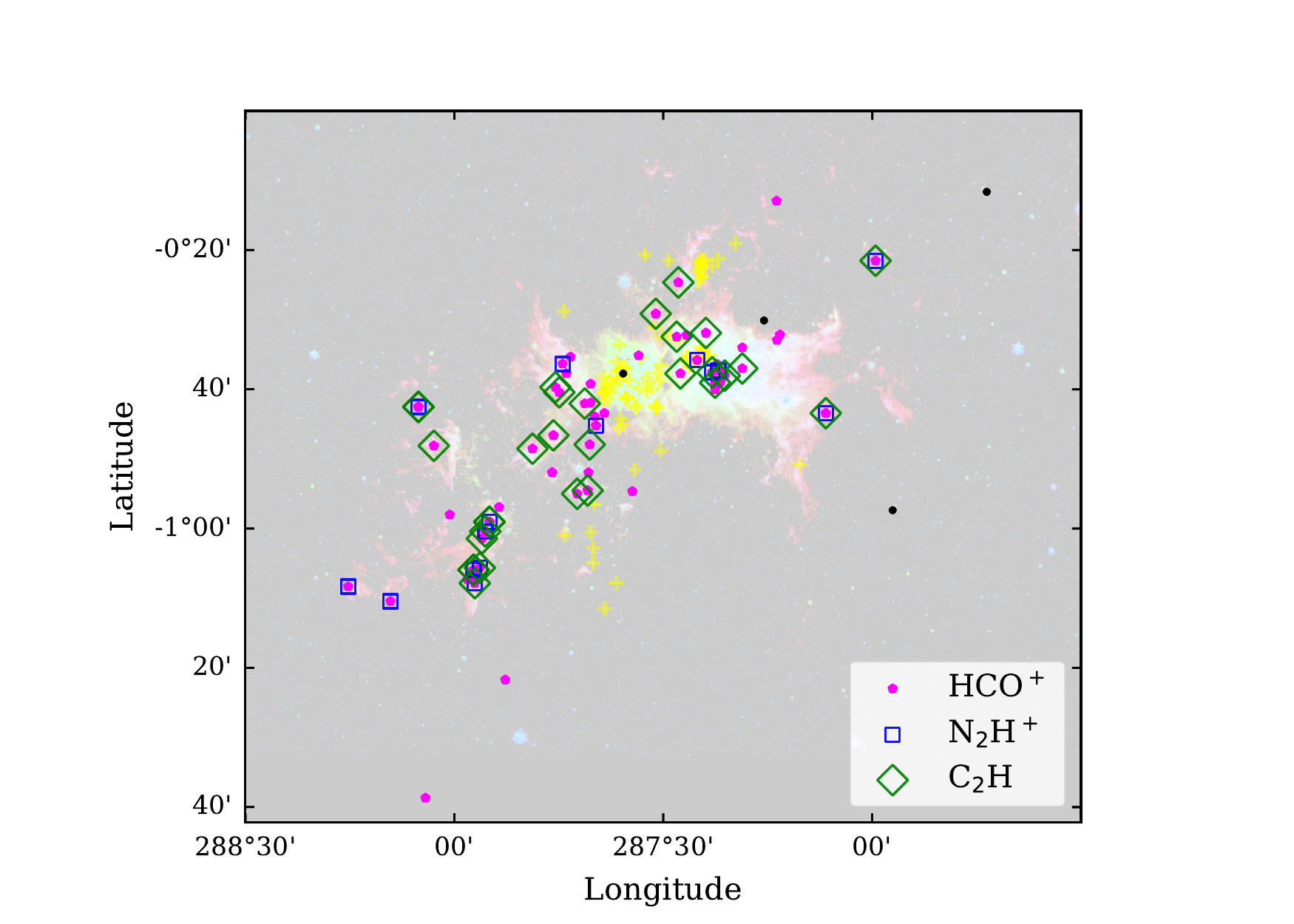}
	\caption{Spatial distribution of the molecules detected towards Carina. Top panel shows the distribution of clumps that present HCO$^+$, HCN and HNC emission. Bottom panel shows the clumps that show emission in N$_2$H$^+$, C$_2$H, in this panel we also included the HCO$^+$ detection to make the comparison between the different molecules easier. In both panels the background shows the same three colour composite as in Figure \ref{Figsources}, the yellow crosses show the position of the high-mass stars, and the black dots shows the position of the clumps that had no molecular line detections. The central black dot corresponds to $\eta$ Carina, which is expected to not have any molecular detection as this is a main sequence star.}
             \label{Figspatial}%
 \end{figure*}

Figure \ref{Figspatial} shows the five molecules detected toward Carina overlaid on a three colour image made from the \textit{Spitzer} GLIMPSE maps, and the position of the high-mass stars in Carina. For C$_2$H, commonly associated to PDRs \citep[e.g.][]{Fuente-1993}, but also to the earliest, cold pre-stellar stages \citep{Sanhueza-2013}, there is no clear spatial trend. However, most of the detections are in regions that are infrared bright. The few clumps that show detections in N$_2$H$^+$ are mostly located far away from Trumpler 16 (Tr 16), the cluster that contains $\eta$ Carina. The low detection of N$_2$H$^+$ can be explained due to the higher temperature of the gas that surrounds the massive stars \citep[evident in the temperatures we derived and also shown in][]{Rebolledo-2016}. CO is the main destroyer of N$_2$H$^+$ in the gas phase \citep[see][]{Caselli-1999}. Thus, as the temperature increases, the CO depleted into the dust grain evaporates, thus lowering the amount of the N$_2$H$^+$ molecule present in this region. 

For the HNC and HCN molecules, we found that most of the clumps that have no detection in either HNC or HCN are also located very close to Tr 16, with several clumps showing only HCN emission in the vicinities of Tr 16. A higher temperature of the gas in this region could explain the lack of HNC, since it can be destroyed by the reactions HNC + O $\rightarrow$ CO + NH or HNC + H $\rightarrow$ HCN +H, that can occur at the typical high densities observed in this region.

\begin{figure*}
   	\centering
   	\includegraphics[trim={0cm 0cm 0cm 0cm}, clip,width=0.67\textwidth]{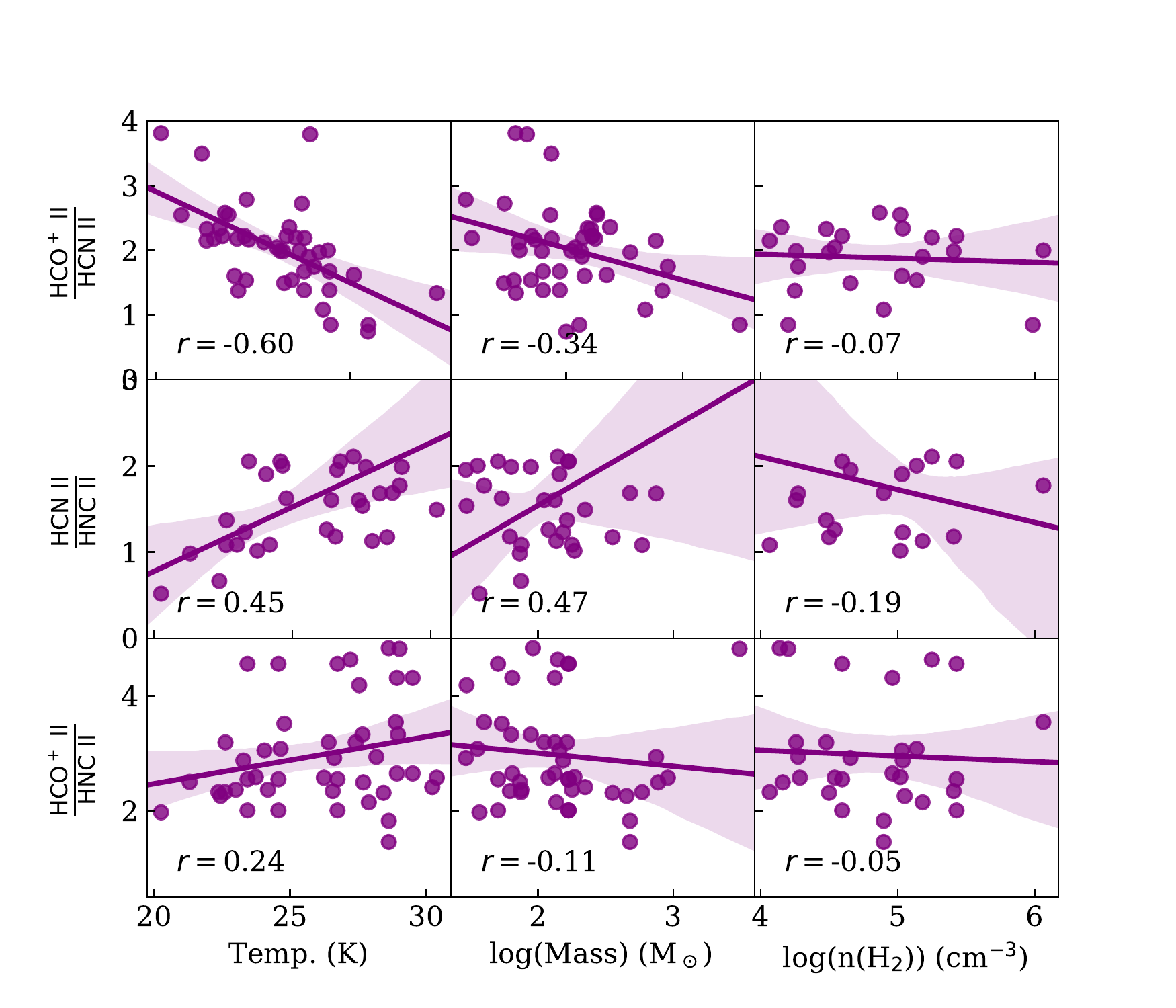}
    
	\caption{Correlation between the integrated intensities of HCO$^+$, HCN and HNC with the temperature, mass and density of the clumps.  In each panel a linear regression to the data is shown with a solid line, and the area of the 95\% confidence interval is shown as the light purple area. The Pearson correlation coefficient (r) is shown in the lower left corner. In these plots, a linear regression with a small confidence area, and a PCC close to 1 or -1 indicates a good correlation between the parameters. The HCN/HNC and HCO$^+$/HCN ratios seems to have a slight correlation with the temperature of the clumps.}
    \label{fig:correlations}%
 \end{figure*}

\begin{figure*}
   	\centering
   	\includegraphics[trim={0cm 0cm 0cm 0cm}, clip,width=0.75\textwidth]{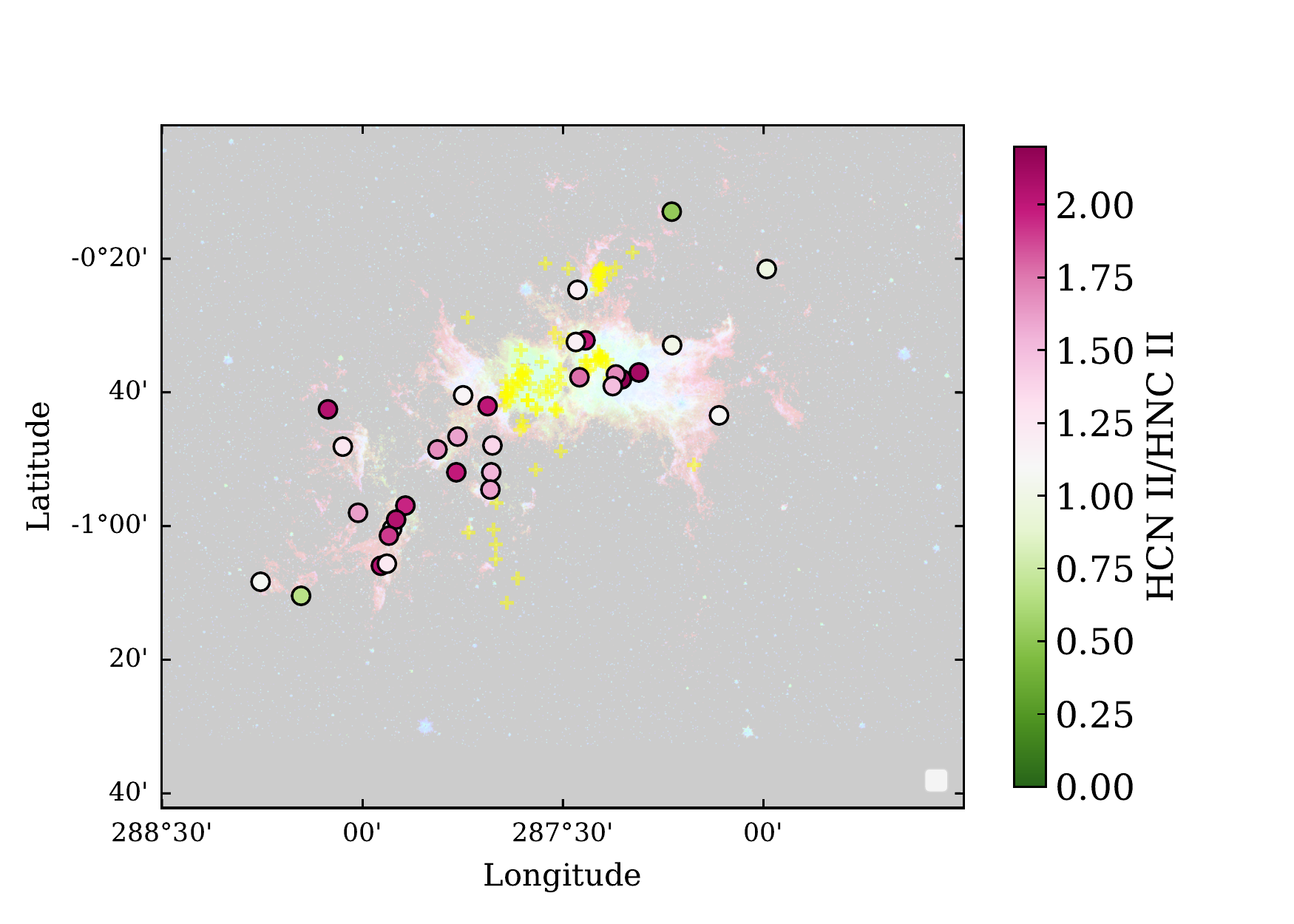}
    
	\caption{Spatial distribution of the ratio between the HCN and HNC integrated intensities. The yellow crosses show the position of the high-mass stars. Colors shows the ratio between the HCN and HNC integrated intensities for the clumps where both molecules were detected.}
    \label{ratioshnchcn}%
 \end{figure*}

The relative abundance ratio of HCN to HNC [X(HCN)/X(HNC)] has been observed to increase as the temperature raises \citep{Goldsmith-1981,Churchwell-1984,Hirota-1998,Tennekes-2006,Hoq-2013,Rathborne-2016}. We assumed that the HNC and HCN emission is optically thin (we cannot verify this as no emission from isotopologues have been measured), and used the ratio of their integrated intensities (II) to determine if there is a trend between these molecules and the clump dust temperatures. To determine any trends, we computed the Pearson correlation coefficient (PCC) between the ratio of the HCN II to the HNC II and the temperature. The PCC is a measure of the linear correlation between two variables, where where 1 and -1 are total positive and negative linear correlation respectively and 0 is no linear correlation. For the clumps in Carina, we found a PCC of 0.45 between the ratio of the HCN II to the HNC II and the temperature, which suggest that there is slight correlation (see Fig. \ref{fig:correlations}). The average HCN II to HNC II ratio is 1.63, and a few clumps have a ratio lower than 1. It is expected that the ratio of HCN II to HNC II be lower than that in PDR regions dominated by X-ray radiation. In Carina most of the X-ray radiation is concentrated near the regions surrounding $\eta$ Carina \citep{Preibisch-2017}, however the clumps that show a low HCN II to HNC II ratio are located outside the region dominated by X-ray radiation, thus the X-ray radiation cannot explain these low ratios (see Fig. \ref{ratioshnchcn}).

The low detection in N$_2$H$^+$ and HNC are therefore consistent with the high temperature of the gas around Tr 16, where the temperatures might be higher than the dust temperatures measured in this region. 

\subsection{Masers and the Carina Nebula}

Masers are prevalent in regions of star formation, particularly from the most common transitions of water, methanol and OH \citep[e.g.][]{Breen-2010,Green-2017,Voronkov-2014,Hattie-2016}. Some masers, like the 6.7-GHz methanol maser, are exclusively associated with high-mass star formation \citep[e.g.][]{Minier-2003,Xu-2008,Breen-2013} and are therefore especially useful in the study of these sources. The Methanol Multibeam Survey \citep[MMB;][]{green+2009} searched the southern Galactic plane for 6.7-GHz methanol masers, detecting 972 sources to a 3-$\sigma$ detection limit of 0.51~Jy within the survey bounds of 186$^{\circ}$$<$$l$$<$60$^{\circ}$, $b$$\pm$2$^{\circ}$. However, despite completely covering the Carina Nebula, the MMB survey failed to detect a single source within it \citep{Green-2012}, which is surprising given the prominent, ongoing star formation \citep[e.g.][]{Smith-2010,Gaczkowski-2013}. In regions of prominent star formation the MMB survey detected as many 17 sources per degree of longitude \citep{Breen-2015} so naively it seems reasonable to expect there to be detectable 6.7-GHz methanol masers in Carina, especially given its close proximity.  

The Carina Nebula has also been completely searched for excited-state OH masers at 6035-MHz as part of the MMB survey \citep{Avison-2016} but no detections were made. This search revealed 127 of these masers across the entire survey field and so their absence in Carina is somewhat less surprising than for the 6.7-GHz methanol masers. Other searches for ground-state OH masers have not covered the entirety of the Nebula, but have also failed to reveal any emission \citep[e.g.][]{Caswell-1987}.

Surveys for water masers have been somewhat more successful, with five sources currently known within the Nebula \citep{Breen-2018,Scalise-1980,Caswell-1989}, even though sensitive searches for water masers have covered far less of Carina than the unsuccessful searches for 6.7-GHz methanol and 6035-MHz excited-state OH masers. \citep{Breen-2018} suggested that this is consistent with there being few young high-mass stars present in Carina (since water masers are somewhat easier to produce and their presence is not as closely tied to the mass of the young stars), a finding also reflected in \citet{Povich-2011} and \citet{Gaczkowski-2013}.

\begin{figure}\vspace{-0.7cm}
    \epsfig{figure=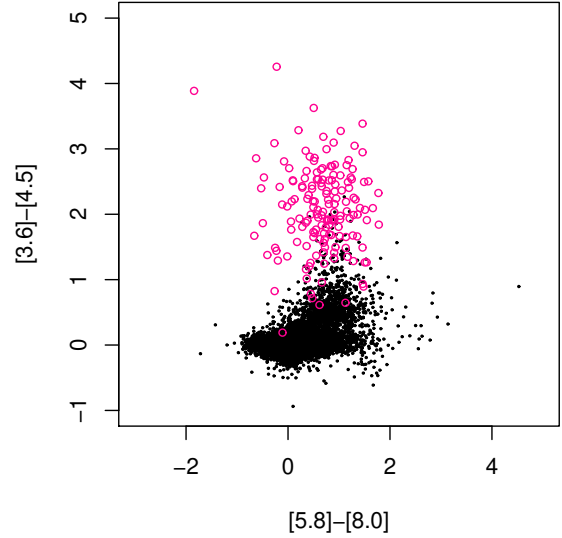,height=8cm, angle=270}
\caption{Colour-colour plot of the GLIMPSE point sources within a one degree radius from G\,287.5−0.6 that have measurements in all four IRAC bands (black dots; 26972 sources). Note that one point source falls outside the x-range of this plot ([5.8]$-$[8.0] = $-$4.878). Pink circles show the GLIMPSE colours of sources associated with 159 6.7-GHz methanol masers from \citet{Breen-2011}.}
\label{fig:glimpse}
\end{figure}

Similarly to \citet{Breen-2018} who compared the GLIMPSE point source colours of water maser associated sources across the Galaxy to the colours of the GLIMPSE point sources located in the Carina Nebula, Fig.~\ref{fig:glimpse} shows the colours of methanol maser associated GLIMPSE point sources in the Galaxy \citep[from][]{Breen-2011} compared to those point sources in Carina. 
There are only 23 sources in Carina that have [3.6]$-$[4.5]$>$1.5, a characteristic held by 81 per cent of the GLIMPSE associated methanol maser sources in Fig.~\ref{fig:glimpse}. For comparison, in a portion of the Galactic plane covering one-quarter of the area and including no comparably prominent regions of star formation (within a 30 arcmin radius of G\,326.5+0), there are 35 GLIMPSE point sources with [3.6]$-$[4.5]$>$1.5 (i.e. over six times as many). This reflects what is shown in Fig.~\ref{fig:glimpse}, while there is some overlap between the methanol maser associated sample and those in Carina, compared to other locations in the Galaxy, there are few sources with a significant 4.5-$\mu$m excess.

\begin{figure*}
   	\centering
   	\includegraphics[trim={0cm 0cm 0cm 0cm}, clip,width=0.48\textwidth]{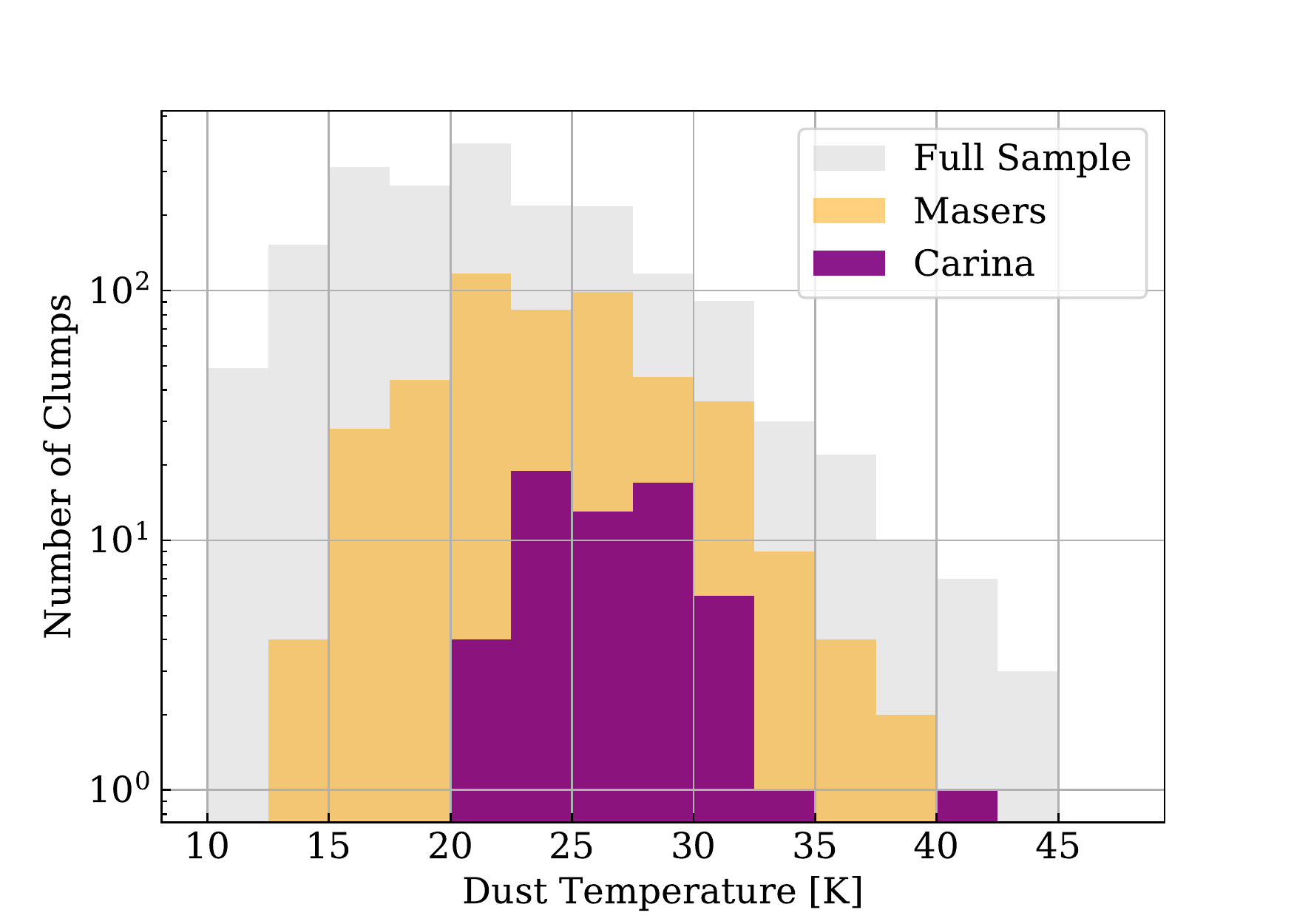}
    \includegraphics[trim={0cm 0cm 0cm 0cm}, clip,width=0.48\textwidth]{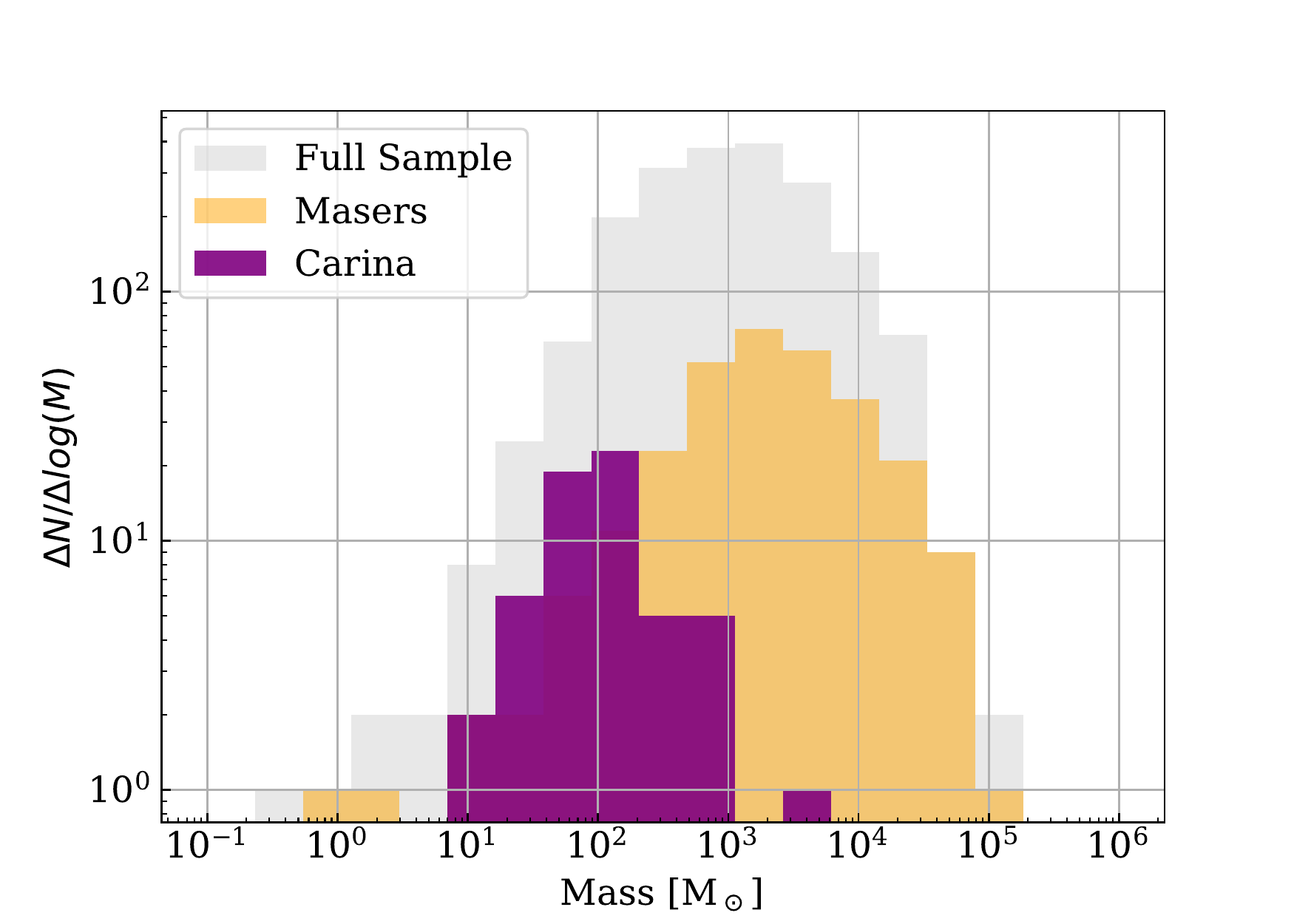}
	\caption{Dust temperature (left) and mass (right) distribution of the total MALT90 sample (grey) compared to those sources in MALT90 that are coincident with a MMB methanol maser (yellow) and those sources located in Carina (purple).}
    \label{fig:mass_temp_masers}%
 \end{figure*}

We have compared the locations of the 6.7-GHz methanol masers from the MMB survey \citep{Green-2012} to the clumps observed in MALT90, using the HPBW of the MALT90 observations as an association threshold. To within this threshold there are 474 of the 3246 MALT90 clumps coincident with one or more 6.7-GHz methanol masers. Fig.~\ref{fig:mass_temp_masers} shows the distributions of the calculated temperatures and masses of the sources in the total MALT90 sample, those MALT90 sources associated with sources from the MMB, and those sources located within the Carina Nebula. As can be seen, the maser-associated sources generally have lower temperatures (mean temperature 24.4$\pm$0.1) and higher masses (mean mass 6000$\pm100$ M$_\odot$) than the majority of sources found in the Carina Nebula (mean temperature of 26.6$\pm$0.1 K, mean mass 214$\pm$5 M$_\odot$). 

The temperatures and masses that we have derived for the maser associated sources are similar to those found in separate studies. \citet{Breen-2018b} found that the far-infrared dust temperatures (calculated in work by \citet{Guzman-2015}) of sources associated with 6.7-GHz sources had mean and median temperatures lower than 25~K. \citet{Urquhart-2013} investigated the mass distribution of ATLASGAL clumps associated with 6.7-GHz methanol masers, finding that the peak of the distribution fell in the 2000 - 3000 M$_{\odot}$ range, which is greater than the mass range of ATLASGAL clumps in Carina, aside from the one clump that has a mass of 3080~M$_{\odot}$ (see Fig.~\ref{fig:mass_temp_masers}). While it is clear that there is overlap between the the temperature and mass distributions of the clumps exhibiting 6.7-GHz methanol masers in other parts of the Galaxy with those clumps located in the Carina Nebula, the properties of the majority of the clumps in Carina tend to be warmer and less massive than those commonly associated with 6.7-GHz methanol masers in other parts of the Galaxy.

We have also compared the molecular line properties of the MALT90 sources associated with 6.7-GHz methanol masers with those sources in Carina. The maser associated MALT90 clumps show slightly higher detection rates for all five lines shown in Figures~\ref{Figdetection} and A1 than the total MALT90 sample, meaning that there is an even larger detection rate discrepancy between the maser associated sources and those in Carina. Furthermore, we find that there is a tendency for the maser associated sources to account for relatively fewer of the weaker MALT90 detections. The line ratio of HCN to HNC has been previously shown to increase with evolution \citep[e.g][]{Hoq-2013,Rathborne-2016}, likely because HNC is more abundant in less evolved clumps. We find that the sources in Carina have line ratios that fall within the range of HCN to HNC ratios of clumps with methanol masers, suggesting that the clumps are not at an evolutionary stage that would preclude associations with methanol maser emission.

\section{Summary}

We performed observations of 16 spectral lines toward 61 clumps in Carina with the Mopra telescope. Observations of the dust continuum emission toward these clumps suggest that their masses are consistent with low and intermediate star forming clumps. Following the empirical relationship of \citet{Larson-2003}, we analyse whether these clumps have sufficient mass to form high-mass stars, finding that only 10\% of them have enough mass to potentially evolve to host high-mass stars. 

The molecular line emission shows very low detection rates compared to the total MALT90 sample of clumps observed in the Galactic Plane. Dividing the clumps into evolutionary stages, following the MALT90 classification, we found that the detection rates are more consistent with those associated with PDRs. However, the detection rates are still low for most of the molecules. 

By subdividing the PDR clumps sample into temperature bins, we found that the N$_2$H$^+$ low detection rate is similar to PDR clumps with high dust temperature clumps (T$_{dust}>35$ K).

We also found that in general the peak antenna temperatures and line width of the molecular line emission are smaller in the Carina region compared to the rest of Galactic clumps.

Comparing the spatial distribution of the clumps within Carina, we found that the clumps that showed N$_2$H$^+$ emission, are usually far away ($>20\arcmin$, 13 pc at a distance of 2.3 kpc) from $\eta$ Carina, and therefore, in regions where the radiation from the star is not enough to drastically increase the dust temperature of the clumps. A similar behavior is seen for HNC, which is not detected in the vicinity of  $\eta$ Carina. These results suggest that the temperature of the gas might be higher than the dust temperature measured in this region. 

To understand the lack of 6.7-GHz maser emission in Carina, we compared the properties of the clumps in Carina to MALT90 clumps  associated 6.7-GHz methanol masers emission. We found that the clumps in Carina are warmer and less massive than those commonly associated with 6.7-GHz methanol masers in other parts of the Galaxy. Moreover, while most of the clumps associated to 6.7-GHz methanol masers in MALT90 have a higher detection rate in the four main molecules HCO$^+$, N$_2$H$^+$, HNC and HCN, the detection rates of these molecules in Carina are very low for N$_2$H$^+$, HNC and HCN. Such discrepancy, can be due to the extreme radiation field produced by $\eta$ Carina on its environment. 

\section*{Acknowledgements}

We thank our referee Anita Richards for valuable comments that have improved considerably the quality of our paper. This research made use of Astropy,\footnote{http://www.astropy.org} a community-developed core Python package for Astronomy \citep{astropy:2013, astropy:2018}.

\onecolumn
\begin{landscape}
\begin{table}
\caption[]{Properties of the dust continuum and molecular line emission. Column 2 shows the peak flux at 870 $\mu$m from LABOCA. Columns 3 to 11 show the peak intensity, velocity of the peak and velocity dispersion obtained from the Gaussian fits of the HCO$^+$, HCN and HNC emission. Some sources have several components along the line of sight. For these, each component has been fitted separately and the values for each fit is given in this table. Full table is available in the online material.}
\begin{tabular}{lr
	S[table-format=-1.2,table-figures-uncertainty=1]
    S[table-format=-1.2,table-figures-uncertainty=1]
    S[table-format=-1.2,table-figures-uncertainty=1]
    S[table-format=-1.2,table-figures-uncertainty=1]
    S[table-format=-1.2,table-figures-uncertainty=1]
    S[table-format=-1.2,table-figures-uncertainty=1]
    S[table-format=-1.2,table-figures-uncertainty=1]
    S[table-format=-1.2,table-figures-uncertainty=1]
    S[table-format=-1.2,table-figures-uncertainty=1]
	rrrrrrrrrrrrrrrrrrrrrc}
\hline\hline
  \multicolumn{1}{c}{Name} & \multicolumn{1}{c}{870 $\mu$m} & \multicolumn{3}{c}{HCO$^+$} & \multicolumn{3}{c}{HCN} & \multicolumn{3}{c}{HNC}   \\
 & \multicolumn{1}{c}{Peak Flux} & \multicolumn{1}{c}{T$_{a}^*$}& \multicolumn{1}{c}{$V_{LSR}$} & \multicolumn{1}{c}{$\sigma$}  & \multicolumn{1}{c}{T$_{a}^*$}& \multicolumn{1}{c}{$V_{LSR}$} & \multicolumn{1}{c}{$\sigma$}  & \multicolumn{1}{c}{T$_{a}^*$}& \multicolumn{1}{c}{$V_{LSR}$} & \multicolumn{1}{c}{$\sigma$}    \\
$\mathrm{}$ & \multicolumn{1}{c}{$\mathrm{Jy/Beam}$} & \multicolumn{1}{c}{$\mathrm{K}$} & \multicolumn{1}{c}{kms$^{-1}$} & \multicolumn{1}{c}{kms$^{-1}$} & \multicolumn{1}{c}{$\mathrm{K}$} & \multicolumn{1}{c}{kms$^{-1}$} & \multicolumn{1}{c}{kms$^{-1}$} & \multicolumn{1}{c}{$\mathrm{K}$} & \multicolumn{1}{c}{kms$^{-1}$} & \multicolumn{1}{c}{kms$^{-1}$} \\ 
\hline 
AGAL286.726$-$00.194     &	1.25 &     &	       &       &&&	\\
AGAL286.951$-$00.956     &	0.74 &     &	       &       &&&	\\
AGAL286.992$-$00.359     &	0.80 &1.03\pm0.03&	-20.62\pm0.03& 0.86\pm0.03&0.58\pm0.02& -20.53\pm0.02& 0.60\pm0.02&0.61\pm0.04 & -20.87\pm0.04 & 0.58\pm0.04	\\
AGAL287.111$-$00.724     &	1.00 &0.70\pm0.03&	-18.82\pm0.05& 1.03\pm0.05	\\
AGAL287.221$-$00.536     &	2.10 &1.41\pm0.04&	-17.70\pm0.04& 1.22\pm0.04&0.59\pm0.02& -17.69\pm0.04& 1.13\pm0.04	\\
 \hline\hline
\end{tabular}\label{table:summary1} 
\end{table}

\begin{table}
\caption[]{Properties of the molecular line emission. Columns 2 to 5 show the excitation temperature, opacity, velocity of the peak and velocity dispersion derived from the hyperfine fit to the N$_2$H$^+$ emission. Columns 6 to 8 show the peak intensity, velocity of the peak and velocity dispersion obtained from the Gaussian fits of the C$_2$H emission. Some sources have several components along the line of sight. For these, each component has been fitted separately and the values for each fit is given in this table. Full table is available in the online material.}
\begin{tabular}{l
	S[table-format=-1.2,table-figures-uncertainty=1]
    S[table-format=-1.2,table-figures-uncertainty=1]
    S[table-format=-1.2,table-figures-uncertainty=1]
    S[table-format=-1.2,table-figures-uncertainty=1]
    S[table-format=-1.2,table-figures-uncertainty=1]
    S[table-format=-1.2,table-figures-uncertainty=1]
    S[table-format=-1.2,table-figures-uncertainty=1]
    S[table-format=-1.2,table-figures-uncertainty=1]
    S[table-format=-1.2,table-figures-uncertainty=1]
	rrrc}
\hline
\hline
Name &  \multicolumn{4}{c}{N$_2$H$^+$} & \multicolumn{3}{c}{C$_2$H}  \\
 &  \multicolumn{1}{c}{T$_{ex}$}&\multicolumn{1}{c}{$\tau$}&\multicolumn{1}{c}{$V_{LSR}$} & \multicolumn{1}{c}{$\sigma$} & \multicolumn{1}{c}{T$_{a}^*$}& \multicolumn{1}{c}{$V_{LSR}$} & \multicolumn{1}{c}{$\sigma$}   \\
 & \multicolumn{1}{c}{$\mathrm{K}$} && \multicolumn{1}{c}{kms$^{-1}$} & \multicolumn{1}{c}{kms$^{-1}$} & \multicolumn{1}{c}{$\mathrm{K}$} & \multicolumn{1}{c}{kms$^{-1}$} & \multicolumn{1}{c}{kms$^{-1}$} \\ 
 \hline
AGAL286.726$-$00.194     &&&&			  	    		   	      	          &           &      	 &  	     \\		 
AGAL286.951$-$00.956     &&&&									  &           &      	  &  	    \\		 
AGAL286.992$-$00.359     &3.02	\pm0.10	&8.76\pm6.89&	-20.74\pm0.04&0.19\pm0.03	  & 0.36 \pm0.04 &  -20.42\pm0.08   & 	0.55\pm0.08 \\ 
AGAL287.111$-$00.724     &&&&									  &           &      	  &  	     \\		 
AGAL287.221$-$00.536     &&&&									  &           &      	  &  	     \\		 
\hline
\hline
\end{tabular}\label{table:summary} 
\end{table}

\begin{table}
\caption{Physical properties the clumps derived from their dust continuum emission. Mass and dust temperature are obtained from the density and temperature mass computed by \citet{Rebolledo-2016}. The profile asymmetry column indicate whether there asymmetry in the line profiles: (1) HCO$^+$ asymmetry, where no other molecule was detected (2) other molecules were detected, but only HCO$^+$ shows asymmetry, and (3) more than one molecule shows asymmetry. Full table is available in the online material.}
\begin{tabular}{
	l
	S[table-format=-1.2,table-figures-uncertainty=1]
    S[table-format=-1.2,table-figures-uncertainty=1]
    S[table-format=-1.2,table-figures-uncertainty=1]
    S[table-format=-1.2,table-figures-uncertainty=1]
	c
    }
\hline
\hline
\multicolumn{1}{c}{Name}                  & \multicolumn{1}{c}{Mass}       & \multicolumn{1}{c}{Dust Temperature} & \multicolumn{1}{c}{Radius} & \multicolumn{1}{c}{Density} & \multicolumn{1}{c}{Profile asymmetry}\\
& \multicolumn{1}{c}{(M$_\odot$)}   & \multicolumn{1}{c}{(K)} & \multicolumn{1}{c}{(pc)} & \multicolumn{1}{c}{10$^4$ (cm$^{-3}$)} \\ \hline
AGAL286.726$-$00.194  & 61\pm6 & 23.4\pm0.5 & 0.07\pm0.01 & 67.2\pm6.3      &  \\
 AGAL286.951$-$00.956 & 33\pm3 & 30.0\pm0.8 & 0.13\pm0.01 & 6.7\pm0.6       &  \\
 AGAL286.992$-$00.359 & 74\pm7 & 21.3\pm0.5 &  &  		        &  \\
 AGAL287.111$-$00.724 & 75\pm7 & 23.0\pm0.5 &  &  		        & 1.0 \\
 AGAL287.221$-$00.536 & 180\pm20 & 23.6\pm0.7 & 0.22\pm0.01 & 7.4\pm0.7     &  \\
 \hline\hline
\end{tabular}\label{tab:target_mass} 
\end{table}

\end{landscape}

\twocolumn

\bibliography{bibliografia}

\end{document}